\documentclass[oldversion]{aa} 
\usepackage{graphicx}          
\usepackage{natbib}

\newcommand{\ampl}{amplification }
\newcommand{\m}{Mg$_\mathrm{0.95}$Fe$_\mathrm{0.05}$SiO$_\mathrm3$ }
\newcommand{\ta}{$\tau_\mathrm{\perp,\mathrm{av}}$}
\newcommand{\td}{$\tau_\mathrm{\perp,\mathrm{field} }$}

\newcommand{\bp}{$\beta$\,Pictoris}

\newcommand{\bpic}{$\beta$\,Pic}
\newcommand{\hsp}{$h_\mathrm{sp}$ }
\newcommand{\rsp}{$r_\mathrm{sp}$ }

\begin{document} 
\title{Collisional dust avalanches in debris discs} 
\author{Anna Grigorieva\inst{1}
  \and P. Artymowicz\inst{1,2}
  \and Ph. Th\'ebault\inst{1,3} }
\institute{Stockholm Observatory, SCFAB, SE-10691 Stockholm, Sweden
\and 
University of Toronto at Scarborough, 1265 Military Trail, 
Toronto, Ontario, M1C~1A4,  Canada
\and
Observatoire de Paris, Section de Meudon,
F-92195 Meudon Principal Cedex, France}
 \offprints{A. Grigorieva} \mail{anja@astro.su.se}
 \date{Received / accepted} \titlerunning{Collisional avalanches in
 dusty discs}                       
\authorrunning{ A. Grigorieva et al.}
\abstract
{We quantitatively investigate how collisional avalanches may develop
in debris discs as the result of
the initial breakup of a planetesimal or comet-like object,
 triggering a collisional chain reaction due to outward
escaping small dust grains.
We use a specifically developed numerical code that follows both the
spatial distribution of the dust grains and 
the evolution of their size-frequency distribution due to collisions.
We investigate how strongly avalanche propagation depends on  
different parameters  (e.g., amount of dust released in the initial breakup,
collisional properties of 
dust grains, and their  distribution in the disc).
Our simulations show that avalanches evolve on timescales of $\sim\!1000$\,years, 
propagating outwards following a spiral-like pattern, and
that their amplitude exponentially depends on the number density of dust grains
in the system. 
{
We estimate a probability for witnessing an avalanche event as a function of 
disc densities, for a gas-free case around an A-type star, and
find that features created by avalanche propagation can
lead to observable asymmetries for  dusty systems with a $\beta$\,Pictoris-like
dust content or higher.  Characteristic observable features include: 
(i) a brightness asymmetry of the two sides for a disc viewed edge-on, and (ii) a  
one-armed open spiral or a lumpy structure  in the case of  face-on orientation. 
A possible system in which avalanche-induced structures might
have been observed is the edge-on seen debris disc around HD\,32297,
which displays a strong luminosity difference between its two sides. 
}
}

\maketitle

\section{Introduction}

Direct imaging of circumstellar discs \citep[e.g.,][]{Heap00, ClampinKrist03, Liu04, Schneider05}
have provided resolved disc morphologies for several systems
(e.g.,  \bpic, HD\,141569A, HD\,100546, HD\,32297) and have shown that
dust distribution is not always smooth and axisymmetric. 
Warps,  spirals, and other types of asymmetries are commonly
observed \citep[e.g.,][ for the \bpic~system]{Kalas95}.
These morphological features can provide hints on important ongoing 
 processes in the discs and improve our
understanding of the evolution of circumstellar discs and of planetary formation.

The usual explanation proposed for most of these asymmetries is the
perturbing influence of an embedded planet. As an example, the 
warp in the \bpic~disc has been  interpreted as induced
by a jovian planet on an inclined orbit \citep{Mouillet97, Au01}. Likewise, 
for annulus-like
discs with sharp inner or outer edges, the most commonly proposed explanation
is  truncation or gap opening due to planets or bound
stellar companions \citep[e.g.,][]{Au04}, although alternative mechanisms such
as gas drag on dust grains within a gas disc of limited extent have also
been proposed \citep{TA01}.
For spiral structures, authors 
have also been speculating on gravitational instabilities \citep{Fuka04},
as well as on a bound stellar companion \citep{Au04}.

The catastrophic breakup of one single large object
releasing a substantial amount of dust fragments could be 
an alternative explanation for some observed asymmetries.  \cite{WyattDent02}
have examined how such collisionally produced bright dust clumps
could be observed in Fomalhaut's debris disc. Likewise, such clumps have
been proposed by \citet{Telesco05} as a possible explanation
for mid-infrared brightness asymmetries in the central \bp~disc,
but only based on preliminary order of magnitude estimates.
More recently, the detailed study of \citet{KenyonBromley05} investigated
the possibility of detecting catastrophic two-body collisions 
in debris discs and found that such a detection would require
the breakup of 100-1000\,km objects.
The common point between these different studies is that they focus
on global luminosity changes due to the debris cloud directly produced
by the shattering events themselves.

In the present paper, we re-examine the consequences
of isolated shattering impacts from a different perspective,
i.e., by considering the collisional evolution of the produced
dust cloud $after$ its release by the shattering event.
The main goal here is to study
one possibly very efficient process, first proposed
by \citet{Art97}, but never quantitatively studied so far,
i.e., the so-called collisional avalanche mechanism.
The basic principle of this process is simple.
After a localized disruptive event, such as the collisional
breakup of a large cometary or planetesimal-like object,
a fraction of the dust then produced is driven out by radiation pressure
on highly eccentric or even unbound orbits.
These grains moving away from the star with significant radial
velocities can breakup or microcrater other particles farther out
in the disc, creating in turn even more small
particles propagating outwards and colliding with other grains. 
Should this collisional chain reaction be efficient enough,
then a significant increase in the number of dust grains
could be achieved. In this case, the consequences
of a single shattering event, in terms of induced dust production, could
strongly exceed that of the sole initially released dust population.
The outward propagation of the dusty grains could then induce observable asymmetric
features in the disc, even if the initially released dust cloud is 
undetectable.

The goal of this work is to perform the  first quantitative study 
of the avalanche process and investigate the morphology of avalanches
in debris discs, under the assumption that dust dynamics is not
controlled by gas \citep{Lagrange00}. 
For this purpose we have created a numerical code, described
in Sect.~\ref{sec:model}, that enables us to simulate the coupled
evolution of dynamics  and size-frequency distribution of dusty grains. 
The results of our simulations, which explore the effect of several parameters
(total mass and radial distribution of  dust in the disc,
mass and size distribution of the planetesimal debris, physical
properties of the grains and the prescription   for collisional
outcome for grain-grain collisions) are presented in 
Sect.~\ref{sec:res}. In Sect.~\ref{sec:obs} we examine
under which  conditions avalanche-induced features might become observable.
We end with a discussion of the probability of witnessing
an avalanche (Sect.~\ref{sec:discussion}) 
and finally a summary (Sect.~\ref{sec:sum}).

\section{Simplified theory of dust avalanches} \label{sec:theory}

A dust avalanche is a chain reaction of outflowing debris impacting disc
particles and creating even more debris accelerated outwards 
by the star's radiation pressure. 
The basic  principle of this  mechanism can be illustrated by a set of
 analytical equations. We present here a simplified theory
of avalanches based on the order-of-magnitude approach of 
\citet{Art97}, firstly for its pedagogical virtues, but also because it
can serve as a reference that
facilitates the understanding of the main results derived from our
extensive numerical exploration.

Let us assume that $N$ particles of  size $s_{gr}$ (radius)
move through  a cloud of dust grains of size $s$  
at a relatively high velocity. Let us further
assume that each collision produces 
a constant number $N_\mathrm{\beta}$ of such  debris,
which are quickly accelerated 
to velocities leading to further destructive collisions.
To derive the total number of debris produced by the avalanche,
we define the optical depth  as
\begin{eqnarray} 
d\tau= n(s) \sigma(s) dl,
\label{eq:tau} 
\end{eqnarray} 
where $n(s)$ is the number density of dust particles of size $s$ 
in the system, 
$\sigma(s) \approx \pi (s+s_\mathrm{gr})^2$ 
is the cross-Sect. for
collisional interaction between grains, and $dl$ is the length measured
along the grain path.
The number of debris produced  in the interval $d\tau$  is then
\begin{eqnarray} 
dN= N N_\mathrm{\beta}\, d\tau. 
\end{eqnarray} 
Integration over the whole path of the grains gives the total number of
debris produced by the avalanche   
\begin{eqnarray} 
N_\mathrm{tot}= N_0{} \exp (N_\mathrm{\beta}  \tau_\mathrm{}), 
\label{eq:N}
\end{eqnarray}
where $N_0$ is the number of outflowing grains 
initially released.

{
In a disc, $\tau$ can be approximated  by the 
optical thickness in the disc midplane,
\begin{eqnarray} 
 \tau_\mathrm{\|}=\int\!\!\int \pi s^2 dn(s) dR,
\label{eq:t}
\end{eqnarray}
where $R$ is the radial cylindrical coordinate. 
We replace $N_\mathrm{\beta}$ by its average value 
$\langle N_\mathrm{\beta} \rangle$  to emphasize   
the fact that in reality $N_\mathrm{\beta}$ depends on the details 
of each collision. 
Equation~\ref{eq:N} then takes the form 
\begin{eqnarray} 
N_\mathrm{tot} \sim N_0{} 
\exp( \langle N_\mathrm{\beta} \rangle \tau_\mathrm{\|}). 
\label{eq:Nsimpl}
\end{eqnarray}
}
This equation gives an estimate of an avalanche 
efficiency in a disc through the total number of grains $N_\mathrm{tot}$
it produces. However, one should  keep in mind that 
the relevance of this set of equations is
limited to global, order of magnitude
estimates. Furthermore, these equations do not give any insight into the 
temporal development and 
spatial structure of a given avalanche. For these crucial issues,
numerical modeling is clearly required.

 \section{The model}
 \label{sec:model}

The number of dust grains in a circumstellar disc is far too large
to follow every grain individually during the calculation; some kind
of statistical approach must therefore be used.
Models of dust disc evolution developed to date
fall into two main categories. 
On the one hand, ``particle in a box''
 models divide the dust grains into statistical bins according to their size and 
 enable us  to
compute the evolution of the size distribution within a 
 given spatially
homogeneous region \citep[e.g.,][]{Th03}.
While it is possible to mimic a spatially inhomogeneous 
system by integrating
a set of coupled particle-in-a-box models, this can become unwieldy in the
absence of strong simplifying symmetries. \citet{KenyonBromley04} use a 
multiannulus code for example, but their model is one-dimensional in space.
On the other hand, direct N-body simulations (treating the dust
as test particles in the potential of a 2 or multi-body system)
are used to accurately follow the spatial evolution of
dynamical structures such as planet induced gaps or resonances
\citep[e.g.,][]{Wyatt03,Au04}. 
In this case, however, the sizes of the dust grains are either not 
taken into account or assumed to be equal.

For the present problem, however, we need to follow both the spatial
distribution of the grains $and$ their size distribution with reasonable 
accuracy.
To do this, we developed a new code in which all grains with similar
parameters (size, chemical composition, spatial coordinates, and velocity)
are represented by a single  { \emph{superparticle}}
(hereafter SP). We follow the dynamical evolution of these SPs 
and compute
the collisional destruction and production of grains as SPs 
pass through
each other. We represent newly created grains as
new SPs. The maximum number of SPs our code can handle is
about one million.

\subsection{Superparticles }

A detailed description of our SP modeling
is given in the appendix. Here we briefly outline
its main characteristics:
A SP is described by the position and velocity of its 
center of mass (which coincides with its geometrical center), 
by its size, shape, and internal density profile,
and by the number of dust grains it contains.
For the present work all SPs are treated as cylinders and their geometrical
centers are constrained to lie in the midplane of the disc. 
The cylinders have constant radii $r_\mathrm{sp}$
and variable heights $h_\mathrm{sp}(R)$, where $R$ is the distance
from the star (see Appendix~\ref{app:sp}).
All grains inside a given SP are assumed to have the same physical properties.
We assume that all grains in our simulation are spheres with
identical densities, chemical compositions, and porosities. 
The grains (and thus the SPs) are 
distributed into mass bins separated by a factor 2
logarithmic mass increment (i.e., a factor of 1.26 in size). 

The trajectory of a SP
corresponds to the trajectory of a test particle (with
dynamical properties identical
to the SP's grains) located at the SP's center of mass 
(see Sect.~\ref{sec:tr}). 
SPs can overlap and freely pass
through each other. In this event,
collisional interactions between their respective grain populations
is considered. 
This process is treated as a passage of two clouds of
grains through each other (see Appendix~\ref{app:new_sp}). It results in
the loss, by destructive collisions, of a fraction of the initial
grain populations and the production of smaller collisional fragments.
These newly produced debris are placed
into newly created SPs in accordance with the 
grain sizes and velocities.  
{
In the current version of the code, the centers of all SPs move in the same plane
and the dust distribution is symmetric with respect to this midplane.
However, the SPs representation method could in principle 
 be used to model systems 
with vertical deformations (e.g., warps). 
}

{
The size of a SP is fairly large ($r_\mathrm{sp}=5$\,AU). 
This puts unavoidable constraints on 
the spatial resolution of our simulations and prevents us from 
modeling processes occurring on scales smaller than the SP radius. 
It would, for example, be difficult
to model fine resonant structures induced by 
disc-planet interaction. Moreover, 
the current version of the method with a constant value of 
the SP radius is not applicable to collisional evolution 
in the inner regions ($\la 20$\,AU) of debris discs. 
Although this limitation could be overcome
by introducing a dependence of the size of a SP on the distance 
to the star (e.g.,  $r_\mathrm{sp} \propto R$), 
we have not implemented it in the current version 
of the code, since our main goal here is  to model collisional 
avalanches 
that propagate outwards, inducing observationally significant 
features in the outer ($\ga 100$\,AU) regions of the disc.
}

{
The grains inside a SP do not have 
explicit vertical velocity components. To check the validity of this
assumption, we have performed test runs, for which an artificial vertical 
velocity dispersion term was added to the planar velocity, which
showed no significant departure from the in--plane velocities case.
Note that a vertical velocity component is, however, indirectly
taken into account by the fact that
SP heights increase with distance from the star (see Appendix~\ref{app:new_sp}),
accounting for the geometrical dilution of grain spatial densities.
}

\subsection{SP trajectories} \label{sec:tr}

As has been  mentioned earlier, the trajectory of a SP is identical to
the trajectory of a test particle (with mass, size, and chemical
composition identical to those of the SP's grains)  
located at the SP's center of mass.
Test particles move  
in the gravitational field of a star under the influence of the stellar
radiation force. The equation of motion reads:
\begin{eqnarray} 
\label{eq:F}
 m\frac{d^2\vec{r}}{dt^2}= - \frac{GMm}{r^3}\vec{r}+\vec{F_\mathrm{rad}}+\vec{F_\mathrm{PR}},
\end{eqnarray} 
where $m$ and $\vec{r}$ are the mass and position of the test particle,
$G$ is the gravitational constant, $M$ is
the star mass, and
$\vec{F_\mathrm{rad}}$ and $\vec{F_\mathrm{PR}}$ are 
the radiation pressure   
and Poynting-Robertson
drag, respectively. 
In our simulations we can neglect the Poynting-Robertson
drag, since it acts on a timescale much longer than the time intervals
considered here. 

\begin{figure}[t]
\begin{center}
\includegraphics[width=\columnwidth,clip]{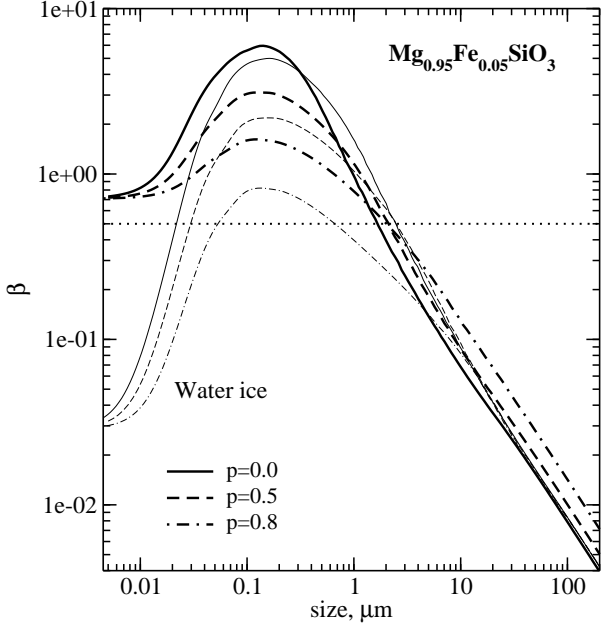}
\end{center}
\caption{Ratio of the radiation pressure force to the gravitational
force  vs. grain size for different grain materials and porosities, $P$, 
 calculated for an A5V (\bp-like) star. The thick lines represent silicate \m. 
The thin lines are used
for water ice. The solid lines are for solid grains, the dashed lines are for 
50\% porous grains,
and the dash-dotted lines are for 80\% porous grains.}
\label{fig:beta}
\end{figure}

The radiation pressure force is expressed  as a function of the
gravitational force  through the radiation pressure coefficient,
$\beta,$ as
\begin{eqnarray} 
\label{eq:Frad}
\vec{F_\mathrm{rad}} = - \beta \vec{F_\mathrm{grav}} = \beta \frac{GMm}{r^3}\vec{r}.
\end{eqnarray} 
The parameter $\beta$ is a function of the stellar luminosity, 
grain size, and optical properties of the grain material \citep{BurnsLamy79}.
We use the  Mie theory code developed  by  \cite{Art88} to calculate
$\beta$ (Fig.~\ref{fig:beta}). 

{
A $7^{th} - 8^{th}$\,order Runge-Kutta method is used for integrating 
test particles trajectories.
Although in the simulations presented in this paper the dynamics of the SPs 
is purely Keplerian, we have decided not to use 
analytical solutions  
since  the Runge-Kutta integrator 
allows for an easy inclusion of any additional gravitational 
(due to planetary or stellar perturbers) or dissipative  forces 
(such as PR and gas drag).

}

\subsection{Initial dusty disc structure}

The SP representation method is used to model the 
initial dusty disc structure. 
The total number of SPs for {\emph{each size bin}}
is chosen so that, at any given
location in the disc, there are at least 2-5
 overlapping SPs to account 
for different dynamical characteristics of grains of this size at
this location. 
Each of these overlapping SPs thus differs from the others by 
its local velocity. To model the initial dust distribution in 
the  disc we use  $\simeq 5 \times 10^4$ SPs (test runs with larger
number of SPs  do not lead to significant changes in the results).
The number density of dust grains at a given location
in the disc is calculated as the sum of the grain densities of the overlapping SPs.

The archetypical, and still by far the best known,
debris disc of $\beta$\,Pictoris is taken as 
a reference system for the initial dusty disc structure. 
In the present study we do not aim
to model this particular system and just adopt its 
global properties for the dust distribution.
Alternative dust distributions are also explored in Sect.~\ref{sec:fp}.
For the dust profile in $\beta$\,Pictoris, we take
the results of \cite{Au01}, who numerically derived the dust distribution
giving the best fit to the resolved scattered light
images as well as the long-wavelength photometric data, as a reference.
We assume here that all grains are produced from parent bodies 
on circular orbits following the best-fit parent body distribution given in 
\cite{Au01}, { where most of the bodies are located within an extended
annulus between 80 and 120\,AU, with a depletion in the inner $<$50\,AU region
and a sharp drop of the density distribution outside 120\,AU
\citep[see for example Fig.~1 of][]{Th05}.}
Grains with small $\beta$
have almost the same orbits as their parent bodies (the biggest grains),
while smaller grains (i.e., with higher $\beta$) have more elliptic orbits 
depending on their
$\beta$ value. The initial number of grains as a function of their
size follows a classical single power law size-frequency distribution 
\begin{eqnarray} 
 dn=N_\mathrm{0s} (s/s_0)^{-p_\mathrm{0s}}ds
 \label{eq:dn}
\end{eqnarray}  
where the power-law coefficient, $p_\mathrm{0s}=3.5$,
corresponds to an { idealized} collisionally  evolved system \citep{Dohnanyi69}.
The minimum size $s_\mathrm{min}$
for the disc grains is given  by the { radiation pressure}
blow-out cutoff and
corresponds to $\simeq 2\,\mu$m for the compact grains considered in
the nominal case. The maximum size $s_\mathrm{max}$ for the disc grains is
taken to be
$1$\,cm. Runs with an order of magnitude higher maximum grain size give very
similar results while being more computationally demanding. At the same time 
we cannot lower 
$s_\mathrm{max}$ since millimeter particles make up a few 
percent of the disc's optical
thickness and their contribution starts to be significant
for the avalanche development. 

The vertical structure of the disc is
 expressed in terms of the vertical 
geometrical optical thickness, $\tau_\mathrm{\perp}$, per unit
length, $z$, as
\begin{eqnarray} 
\label{eq:dtaudz}  
 \frac{d\tau_\mathrm{\bot}}{dz} =C_\mathrm{\tau} \frac{\tau_\mathrm{\bot}}{w}  
 exp \left[ - \left(\frac{|z|}{w}\right)^{p_\mathrm{z}}\right],
\end{eqnarray}   
where $C_\mathrm{\tau}$ is a normalizing constant,  
$\tau_\mathrm{\bot}(R)=\int\!\!\int \pi s^2 dn(R,s) dz$ is
the vertical optical thickness of the disc at distance $R$
from the star, $n(R,s)$ is the number density of dust grains of size $s$, 
$w(R)$ is the disc width, and  $p_\mathrm{z}=0.7$ is a parameter,
that determines the shape of the vertical profile. 
The disc width changes with radius as
\begin{eqnarray} 
\label{eq:w}
 w_\mathrm{}(R)=0.055 R_\mathrm{m}  \left(\frac{R}{R_\mathrm{m}}\right)^{p_w},
\end{eqnarray} 
where $R_\mathrm{m}=117$\,AU and $p_w=0.75$ for most of the runs. (Alternative values of
$p_w$ have been explored in test runs, which have shown that results only weakly depend
on it.)

\subsection{Collisional outcomes} \label{sec:col}

Collisions are the crucial mechanism for the development of the
avalanche phenomenon. The result of a collision, 
in terms of the 
size-frequency distribution of the debris, depends on several parameters:
projectile and target materials and structures, sizes, impact velocities,
and angle of incidence. Since it is   not possible to model every collision in such
detail, we  have to adopt a simplified algorithm.
We assume that the impact energy of colliding bodies, $E_\mathrm{col}$, is equally shared between them.
Laboratory experiments show that this is the case when both bodies are made from
identical material regardless of their sizes  \citep{Ryan91}. $E_\mathrm{col}$ is:
\begin{eqnarray*} 
 E_\mathrm{col}=\frac{M_1 M_2 v_\mathrm{rel}^2 }{ 2(M_1+M_2)},
\end{eqnarray*} 
where  $v_\mathrm{rel}$ and $M_\mathrm{1,2}$ are the relative 
velocity and the masses of the colliding bodies. The
 relative velocity between grains is derived from 
relative velocities between SPs after compensating for the artificial
Keplerian shear induced by the SP finite radius.

Collision outcomes are traditionally  divided into two classes: 
(1) catastrophic fragmentation, when the largest remaining fragment, $M_\mathrm{lf}$,
is less than  half of the parent body mass, $M$,
and (2) cratering, when $M_\mathrm{lf} > 0.5M $. The energy per unit mass that is 
needed to get  $M_\mathrm{lf} = 0.5M $ is called the threshold specific energy $Q^*$.
If the specific energy $Q=0.5E_\mathrm{col}/M$ received by a body is more than
$Q^*$, then the collision leads to  catastrophic breakup, whereas cratering
occurs if $Q < Q^*$ \citep{Fu77, PF93, Benz99}. $Q^*$ is a function of size for
which we adopt a classical power law dependence \citep[e.g.,][]{RyanMelosh98, HH99}.
The collisional response of the small objects considered in the present work
falls into the so-called strength regime, where the target's internal strength
is the dominant factor, for which
\begin{equation}
Q^*=Q^*_0(s/s_0)^{-p_Q},  
\label{eq:Qth1}
\end{equation}  
where $Q^*_0 $ corresponds to the value of the threshold energy for  size $s_0$.
In this regime, $Q^*$ decreases with size.
\citet{HH90} and \citet{RyanMelosh98}  present a wide range of values for $p_Q$.
We cannot directly apply their result because  we
are dealing with much smaller sizes.  As pointed out 
in \citet{Tielens94} ``for
submicron-sized bodies, cracks play little role, and the strength of a material
approaches the ultimate yield strength of the material'', which corresponds to
$Q^*=2\times 10^8 $\,ergs/g  \citep[][ and references therein]{Tielens94}. 
The threshold energy size-dependence most probably has a knee 
in the micron-submillimeter size range, but since we do not have any information about 
the slope change we decided not to introduce two additional unknown parameters into the
code, but simply to adopt a slightly shallower slope, $p_Q=0.2$, for our calculations.

To account for the effect of different incidence angles, we
correct the value of $Q^*$ by a correction factor $x_\mathrm{cr}$
corresponding to an average over all incidence angles 
\begin{eqnarray*} 
Q^*=Q^*_\mathrm{head\,on}/x_\mathrm{cr},
\end{eqnarray*} 
where $x_\mathrm{cr}=0.327$ \citep{PF93}. 
For both catastrophic shattering and cratering  prescription,
 we use the approach and
the algorithm  presented  in \citet{PF93}. However, we
 refine this model
by assuming that the fragment mass distribution  produced follows
a broken power-law instead of a single-index one: 

\begin{eqnarray}
 dn=N\, dm\left\{ \begin{array}{ll}
  \left( {m}/{m_\mathrm{s}}\right)^{-q_1} &\mbox{if  $m < m_\mathrm{s}$} \\
  \left( {m}/{m_\mathrm{s}}\right)^{-q_2} &\mbox{if  $m \geq
  m_\mathrm{s}$.}
 \end{array} \right.
\label{eq:dn_qi} 
\end{eqnarray}

Such a change of slope between the
small and large fragments, always corresponding to a flattening of the slope
in the small particle range, is indeed supported by experimental results
\citep[e.g.,][]{DavisRyan90}. 
For the choice of values for the power-law
indexes ($q_1, q_2$) and the transition mass $m_\mathrm{s}$ for the slope change,
we take the 
experimental studies of \citet{DavisRyan90} as a reference and explore values within
the range of possible values obtained by these authors.
The minimum fragment size is assumed to be $0.1\,\mu$m, 
unless otherwise explicitly
specified. 

{
In our calculations we do not consider changes in the orbital 
parameters of the colliding bodies (i.e., SP), since 
this effect is not important for the present study. 
There are 2 reasons for this: 
(i) the  lifetime of an avalanche 
(typically $\sim\!10^3$ years) is  very 
short from the point of view of the global disc evolution, 
thus we can neglect any changes in the disc dynamics 
caused by mutual collisions 
between the disc particles (i.e., ``field SPs'' in our simulations);
(ii) the dynamics for the majority of the avalanche SPs are controlled 
by the radiation pressure. Their orbital parameters are thus determined 
mostly by their $\beta$ values and only weakly depend on the velocities
at which these SPs are born (as is verified in Sect. 4.2.1
for the first generation of avalanche grains).
In this respect, taking the velocity 
of the center of mass of the colliding grains as the initial velocity for 
the produced debris is a good approximation within the frame
 of our simulations.   

} 

\subsection{Initial planetesimal breakup} \label{sec:plb}

As previously mentioned, we assume that the initial source of
the collisional avalanche is the breakup of a large, at
least kilometer-sized object. 
{ We do not perform a simulation 
of the initial shattering event itself, 
but implement a simple parametric prescription
for the  dust released in the breakup. In most runs, 
we consider a ``nominal'' case, in which $M_0=10^{20}$\,g 
of dust is released in the $0.1\,\mu$m to $1$\,cm range
at $R_\mathrm{0}=20$\,AU  from the star, unless otherwise 
explicitly specified. 
It should be noted that the released dust mass $M_0$ is the only
relevant parameter for our simulations. 
In this respect, the exact process leading to the initial release is
not crucial.
However, when it comes to estimating the probability for such
a dust-release event to occur (as will be done in Sect.~\ref{sec:Pobs}),
one has to consider the mass $M_\mathrm{PB}$ of the parent
body whose shattering produces a mass $M_0$ of dust. 
The ratio $M_0/M_\mathrm{PB}$ is obviously $<1$, but strongly depends
on several poorly constrained parameters, mainly related to the physics
of the shattering event. 
For an idealized case when 
the largest fragment produced has mass
$M_{lf}=0.5M_\mathrm{PB}$ and smaller fragments follow the Dohnanyi
``equilibrium'' size distribution  ($dn  \propto s^{-3.5}ds$),
one gets $M_\mathrm{PB}\simeq 10^{24}$g. However, 
laboratory and numerical studies
as well as observations of asteroid families 
\citep[e.g.,][]{DavisRyan90,Tanga99}
all point towards smaller $M_{lf}$ and
steeper size distributions for highly disruptive
impacts of large objects, with indexes typically in the $-3.7$ to $-4$ range
for the largest $\geq 0.01M_{lf}$ fragments and closer to $-3.5$
for the smallest ones. Using for example the fragmentation prescription for
large objects of \citet{Th03}, we determine that for a typical shattering 
at 1\,km/s,
$M_\mathrm{PB}\simeq 10^{21}$\,g$=10M_{0}$, which corresponds to an object
 of radius $\sim 40$\,km. 
We shall thus assume a nominal $M_0/M_\mathrm{PB}$ ratio of 0.1 for the discussion
in Sect.~\ref{sec:Pobs}.
For the size spectrum of the dust particles released in the
$0.1\,\mu$m to $1$\,cm range, we assume a single power
law (Eq.~\ref{eq:dn} with $p_{s0}=3.5$) for our
nominal case.
}
The dependence of the results on  $M_0$, $p_{s0}$, and other 
parameters related to the planetesimal debris is 
explored in Sect.~\ref{sec:c}.

\section{Results}
\label{sec:res}

For the sake of the readability of the results, it is convenient to
divide the system into two populations: 1) the {\bf\emph{avalanche
particles}},
representing all bodies initially released by the 
planetesimal breakup plus
all grains later created by
collisions between the avalanche particles and the disc material, and
2) the {\bf\emph{field particles}}, i.e., the population of grains in the disc 
unaffected by the avalanche mechanism.
To quantify the magnitude of an avalanche
we introduce the  {\bf\emph{{area amplification factor}}}, 
$ F_\mathrm{}$,
which is the ratio of the total cross-sectional 
area of the avalanche grains, within
500\,AU from the star, to the initial cross-sectional area 
of planetesimal debris
released. The maximum value $F_\mathrm{max}$ reached
by the amplification factor while the avalanche is propagating
is used to measure the amplitude of a given avalanche and
to compare avalanches obtained for different initial
conditions. Time is expressed in orbital periods at 20\,AU ($\sim$\,70\,yr),
unless otherwise explicitly specified.
Table~\ref{tab:nom} summarizes the set of initial parameters
chosen for our ``nominal'' case. 
All free parameters of the simulations are then explored
in separate runs.

\subsection{Nominal case (NC)}

\begin{table}
\caption{Main model parameters for the nominal case.} 
\begin{tabular}{ll}
\hline\hline
\multicolumn{1}{c}{\bf{Grains:}}\\ \cline{1-1}
material & \m \\ 
porosity  & compact grains (P=0)   \\ 
grain density & 3.5\,g.cm$^{-3}$ \\
\cline{1-1}
\multicolumn{1}{c}{\bf{Disc (``field'' population):}}\\ \cline{1-1}
minimum size  & $s_\mathrm{min}=2\,\mu$m\\  
maximum size &$s_\mathrm{max}=1$\,cm\\
radial distribution & \citet{Au01} \\
optical thickness along radius \\
in the midplane &	$\tau_\mathrm{\|}=0.022$ \\
disc extension & $[20, 500]$\,AU \\
\cline{1-1}
\multicolumn{1}{c}{\bf{Initial planetesimal debris:}}\\ \cline{1-1}
minimum size & $s_\mathrm{min,pl}=0.1\,\mu$m\\ 
maximum size &$s_\mathrm{max,pl}=1$\,cm\\
size distribution (see  Eq.~\ref{eq:dn})  & $p_\mathrm{0,pl}=-3.5 $ \\
initial mass of dust released  & $M_0=10^{20}$\,g \\
distance from the star for \\
the planetesimal breakup &  $R_0=20$\,AU \\
initial velocity of \\
the center of the mass &  $v_0=1.1v_\mathrm{kep} $ \\
\cline{1-1}
\multicolumn{1}{c}{\bf{Collisional prescription:}}\\ \cline{1-1}
threshold energy, $s_0=1$\,cm & $Q^*_0=10^7$\,erg.g$^{-1}$ \\ 
power-law index (Eq.~\ref{eq:Qth1}) & $p_Q=-0.2 $   \\ 
\multicolumn{2}{l}{\bf{size distribution of debris (see  Eq.~\ref{eq:dn_qi}):} }\\
{~~~~}minimum size & $s_\mathrm{min,col}=0.1\,\mu$m\\ 
{~~~~}power-law indexes for $m<m_s$ & $q_1=1.5$ \\
{~~~~}power-law indexes for $m \geq m_s$ &$q_2=1.83$ \\      
{~~~~}position of the slope change  &  $m_s=M_\mathrm{lf}/3$ \\
\hline
\end{tabular}
\label{tab:nom}

\end{table}

\begin{figure*}
\includegraphics[width=0.65\columnwidth,angle=-90]{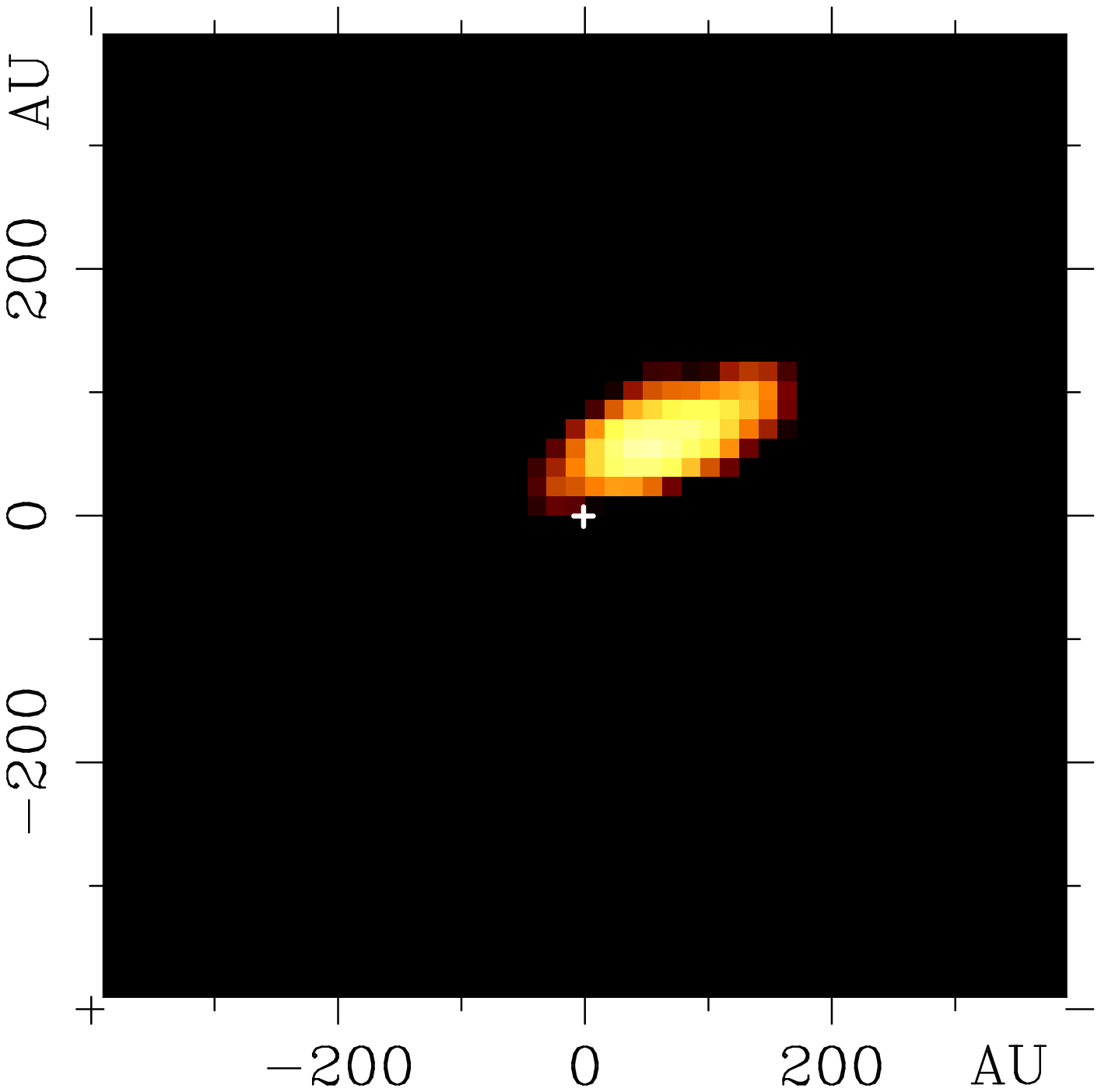}
\includegraphics[width=0.65\columnwidth,angle=-90]{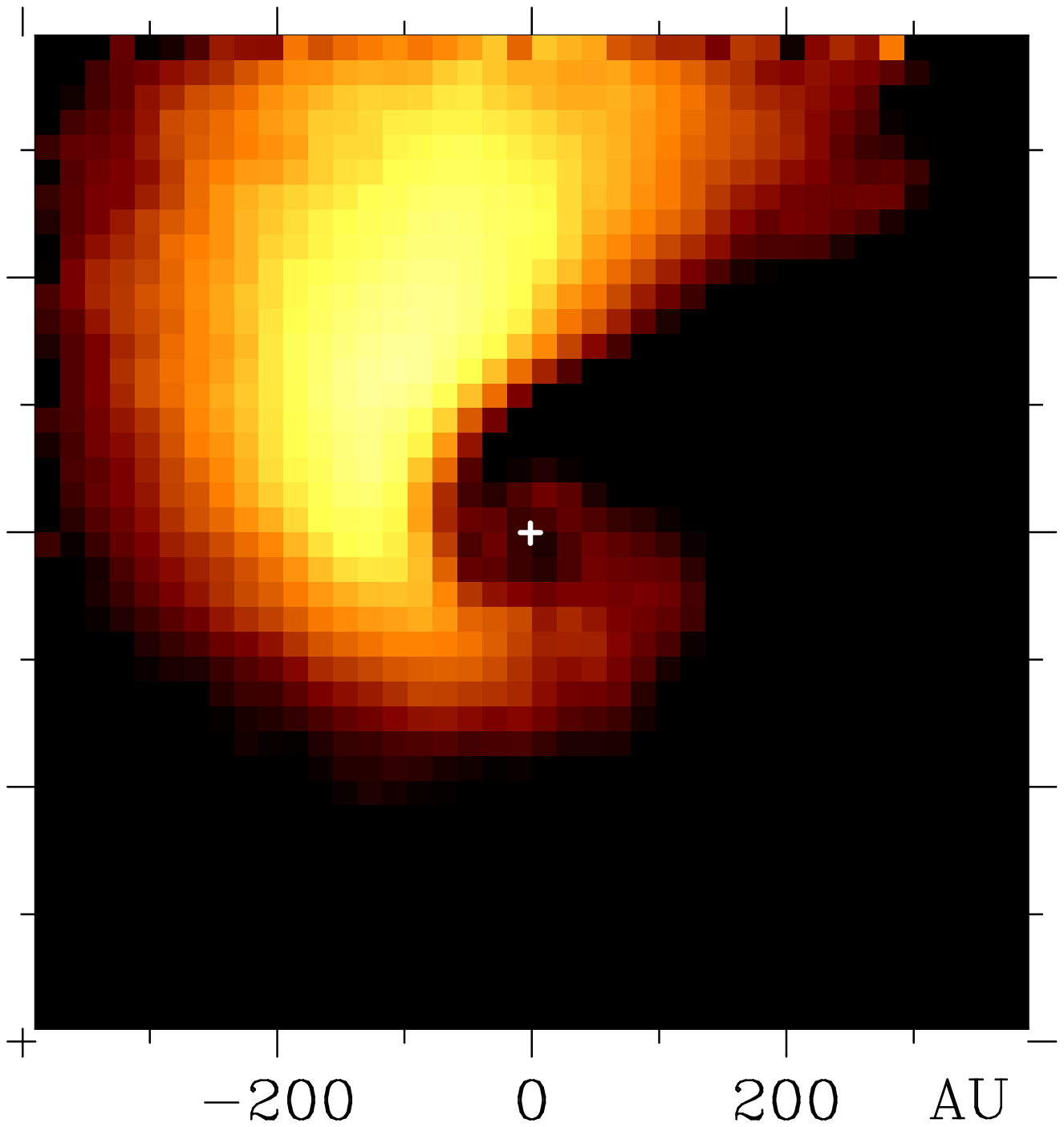}
\includegraphics[width=0.65\columnwidth,angle=-90]{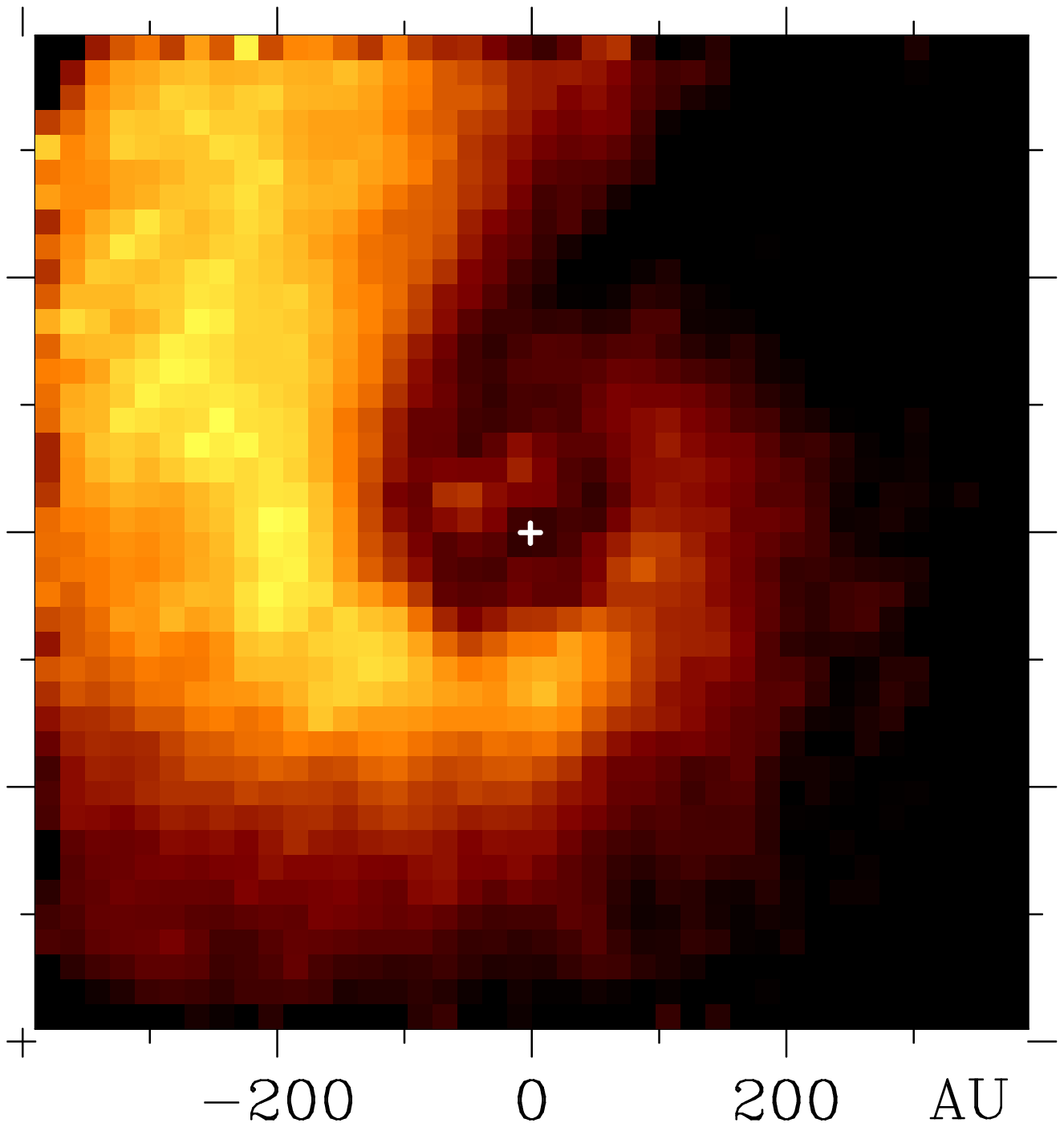}
\includegraphics[width=0.65\columnwidth,angle=-90]{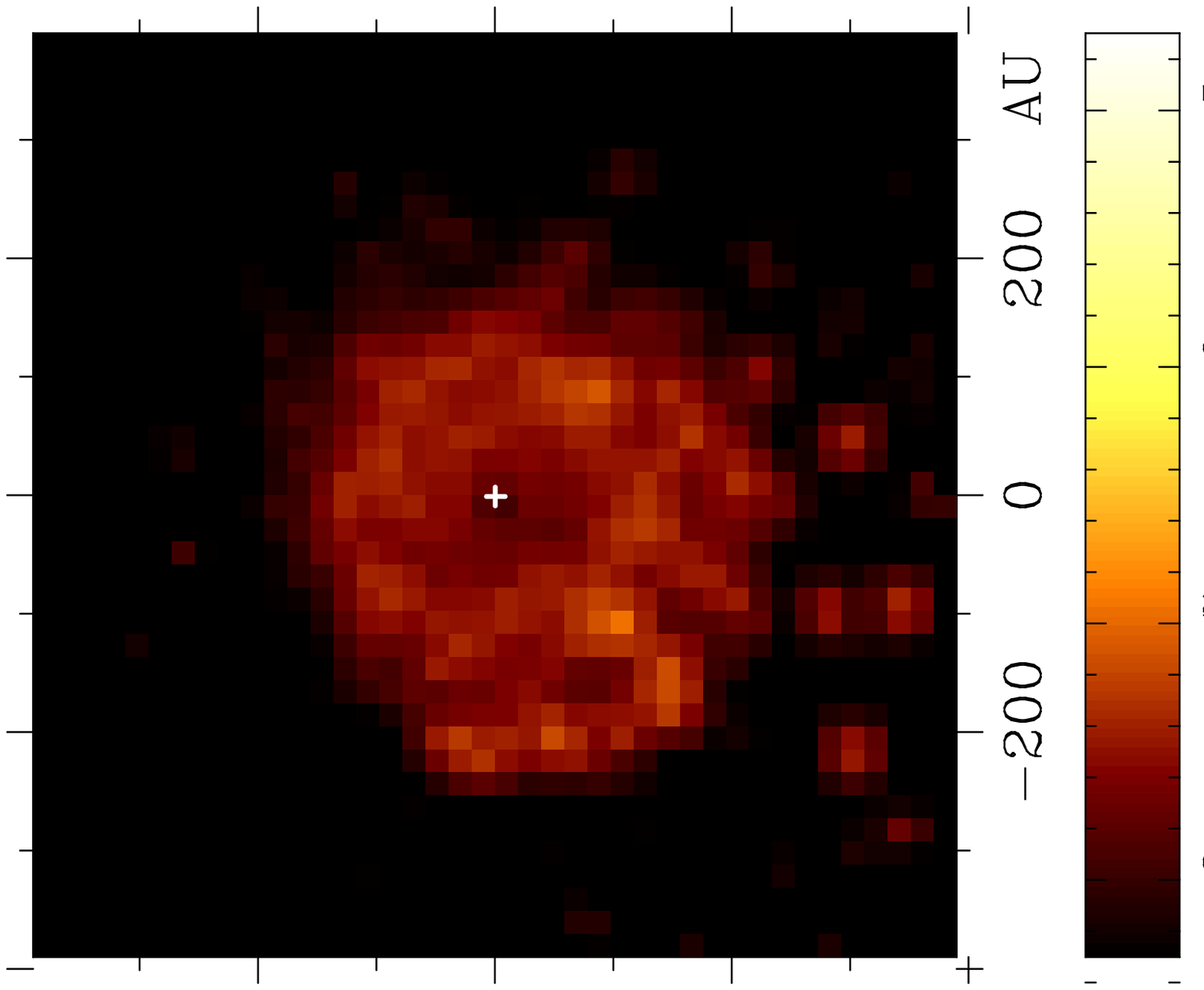}
\caption[]{Nominal case. Color-coded maps (log-scale) of  the 
vertical optical thickness of avalanche grains, \ta , 
at different stages of the avalanche evolution  (t=0.6, 5,
10, 40 orbital periods at 20\,AU). The planetesimal debris
are released at $t=0$ at 20\,AU from the star.
{\emph{Field particles are not
included in the plots.} The position of the star is marked by the 
white cross.}
}
\label{fig:av_disk}
\end{figure*}

\begin{figure}[t] 
\begin{center}
\includegraphics[width=\columnwidth,clip]{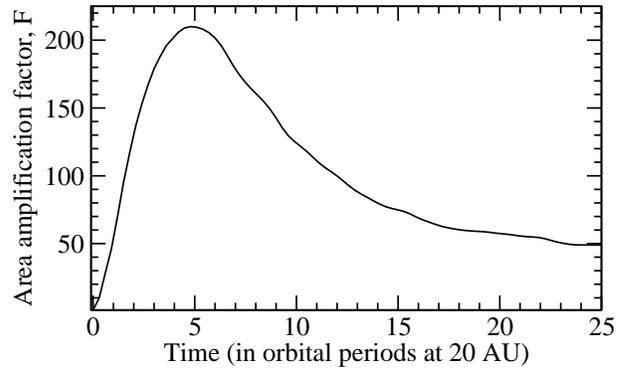}
\end{center}
\caption{Time evolution of the 
cross-sectional area amplification factor (the ratio of the total cross-sectional area 
of the avalanche grains within 500\,AU to its initial value at $t=0$).
Initial increase is due to dust production by outflowing 
planetesimal debris colliding with the
disc material. When the  grain removal (due to star radiation pressure)
rate exceeds the grain production, the value of F
drops (see text for more details).
 }
\label{fig:Ftime}
\end{figure}
   
\begin{figure}[t]
\begin{center}
\includegraphics[width=\columnwidth,clip]{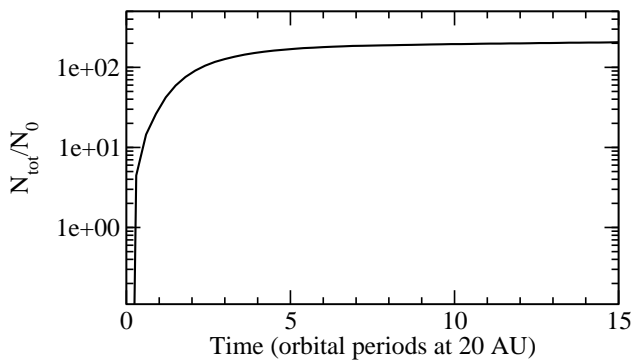}
\end{center}
\caption{Ratio of the total number of grains produced 
by the avalanche 
 by the time $t$, $N_\mathrm{tot}$, 
 to the initial number of released planetesimal debris, $N_0$. }
\label{fig:Ngrtot}
\end{figure}

Figure~\ref{fig:av_disk} shows the temporal evolution of an avalanche,
for the nominal case, in terms of the vertical optical thickness,
\ta, of the avalanche particles ($\tau_\mathrm{\perp, av}=
\int\!\!\int \pi s^2 dn_\mathrm{av}(s) dz$).
As expected, the first stages correspond to
a fast development and multiplication of the avalanche grains. 
In this early expansion
phase the surface density is
dominated by the smallest high-$\beta$ ($\ga 0.5$) particles, which contribute to
$\simeq 85 \%$ of \ta. The maximum value of the amplification
factor is
$F_\mathrm{max}=210 $ and is reached at $t \simeq 5$ ($\simeq 350$\,yrs, 
see Fig.~\ref{fig:Ftime}).
After that, the loss
of small grains on unbound orbits dominates over the collisional production of new
dust particles, and the avalanche begins to fade. In these later stages,
the total cross sectional area of the avalanche grains (within $500$\,AU)
 is increasingly dominated by the bigger grains on bound orbits.
It is important to point out that the timescale for the avalanche propagation 
is short in comparison with orbital periods in the outer part of the disc
(e.g., only $\sim 1/5$ of the orbital period at 200\,AU).

The amplification achieved by the avalanche mechanism is  impressive, i.e.,
an increase in grain cross-sectional surface density by
two orders of magnitude compared to the particles initially released by
the planetesimal breakup (Fig.~\ref{fig:Ftime}). However, absolute
values of \ta~are still very small compared to those of the field
particles, with \ta/\td~never exceeding $10^{-2}$
(see Sect.~\ref{sec:obs} for a more detailed discussion of this crucial parameter).   

To compare the results of our simulation with the 
simplified theory of Sect.~\ref{sec:theory},
we plot the ratio of the total 
number of grains $N_\mathrm{tot}$
produced by the avalanche until time $t$ to the 
initial number $N_\mathrm{0}$ of released planetesimal debris 
(Fig.~\ref{fig:Ngrtot}).
As can be clearly seen,  $N_\mathrm{tot}/N_\mathrm{0}$ quickly reaches
a plateau, and 
we take  $N_\mathrm{tot}/N_\mathrm{0} \simeq 200$ at
$t=15$  as a reference value. Plugging values for the average
number of particles produced by each grain-grain collision,
 $\langle N_\mathrm{\beta} \rangle \approx 150$, 
 and $\tau_\mathrm{\|}=0.022$
 into Eq.~\ref{eq:Nsimpl}, we get $N_\mathrm{tot, theory}/N_\mathrm{0} 
 \approx 30 $, i.e., a factor of $\sim$\,7 difference with the result 
 of our simulation. This is mainly due to the fact that 
 $\tau_\mathrm{\|}$ underestimates the real value of $\tau$, firstly 
because the real path of a grain is curved rather 
than parallel to a disc radius,
and secondly because in  $\tau_\mathrm{\|}$ 
the size of the avalanche grains $s_\mathrm{gr}$ is neglected. 
From our simulations, we were able to estimate the discrepancy 
between $\tau$ and $\tau_\mathrm{\|}$ to be roughly of
a factor of 1.6. We thus get
\begin{equation} 
\frac{N_\mathrm{tot}}{N_\mathrm{0}}\approx  
\mbox{e}^{1.6\tau_\mathrm{\|}\langle N_\mathrm{\beta}\rangle} 
\simeq200,
\label{NN}
\end{equation}
a value close to the numerical results.

It is also interesting to link $N_\mathrm{tot}/N_0$ to the 
amplification factor
parameter $F_\mathrm{max}$.
By definition, 
\begin{equation}
F(t)=\frac{\int s^2 dN_\mathrm{in}(s,t)}{\int s^2 dN_0(s)}=
\frac{N_\mathrm{in}(t) \langle s(t)^2\rangle}{N_0 \langle s_0^2\rangle},
\label{eq:Fmax}
\end{equation}  
where $N_\mathrm{in}(s,t)$ is the total number of avalanche grains 
inside 500\,AU at  
time $t$, and $ \langle s(t)^2\rangle $ and  $\langle s_0^2\rangle $ are
the averaged cross-sectional areas of the avalanche grains at time $t$ and of
the initially released planetesimal debris, respectively.  
Since the avalanche's dust production comes mainly from a rather short
peak of activity (see Figs.~\ref{fig:Ftime} and \ref{fig:Ngrtot}),  
$N_\mathrm{tot}$ is close to $N_\mathrm{in}(t_*)$, where $t_*$ is the time at which 
$N_\mathrm{in}$ reaches its maximum.
A numerical check showed that indeed 
$N_\mathrm{in}(t_*) \approx 2/3 N_\mathrm{tot}$, so that

\begin{equation}
F_\mathrm{max} \approx \frac{2}{3}\,\frac{N_\mathrm{tot}}{N_0} \frac{\langle s(t_*)^2\rangle }{\langle s_0^2\rangle }
= \frac{2}{3} \frac{\langle s(t_*)^2\rangle }{\langle s_0^2\rangle }
~ \mbox{e}^{1.6 \langle N_\mathrm{\beta} \rangle \tau_{\|}} .
\label{eq:FN}
\end{equation}  
The validity of this relation is easily numerically verified, with
$2/3 N_\mathrm{tot}/N_0 \times \langle s(t_*)^2 \rangle/ \langle s_0^2 \rangle \simeq 200$,
a value that is indeed relatively close to the $F_\mathrm{max}$
value obtained in the simulation.

\subsection{Dependence of $F_\mathrm{max}$ 
on the initial planetesimal debris parameters} 
\label{sec:c}

\subsubsection{Initial mass \& velocity of the planetesimal dust cloud}
\label{sec:mc}

The initial mass $M_0$ of dust released
has been explored as a free parameter. The simulations show that the 
maximum \ampl factor, $F_\mathrm{max}$, does not vary with $M_0$, at least in the
$10^{12}$\,g--$10^{21}$\,g range, a result that  is in good agreement with
Eqs.~\ref{eq:FN} and \ref{eq:N}.
Likewise, $F_\mathrm{max}$ does not change much when varying the initial 
speed $v_\mathrm{0}$ of the center of mass of the planetesimal dust cloud.
There is only a 20~\% increase of $F_\mathrm{max}$ when $v_\mathrm{0}$ is increased from 
$v_\mathrm{kep}$ to $1.41v_\mathrm{kep}$.
This weak dependence on the
initial velocity of the debris confirms the fact that  avalanches are driven mostly
by the smallest particles, which are accelerated to high speeds 
weakly correlated
to the initial release velocity.

\subsubsection{Size distribution of planetesimal debris}
\label{sec:sminc}

\begin{figure}[t] 
\begin{center}
\includegraphics[width=\columnwidth,clip]{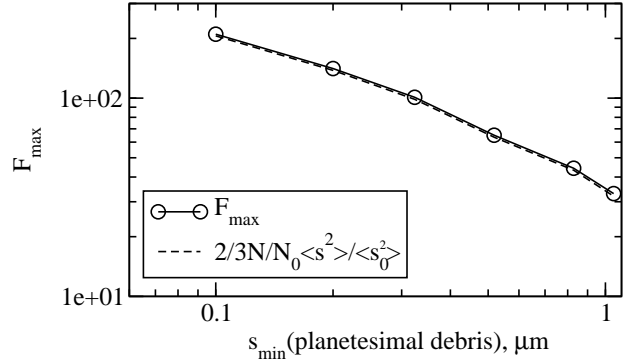}
\end{center}
\caption{Maximum amplification factor as a function  of the 
minimum size assumed for the initial planetesimal debris. The power-law index for the size
distribution is equal to its nominal value $p_\mathrm{0,pl}=3.5$.
For smaller $p_\mathrm{0,pl}$ the variation with $s_\mathrm{min,c}$ is less pronounced.
}
\label{fig:fm_sminc}
\end{figure}

For the nominal case we choose $s_\mathrm{min,pl}=0.1\,\mu$m. This value
is compatible with the lower limit for the size of the
interplanetary dust particles \citep{Fr82}.
It is also in good agreement with the size distribution deduced 
from studies of
cometary comas that show that the smallest particles are
about $0.08-0.28\,\mu$m in diameter \citep{Kolokolova01}. \citet{McDonnell91}
observed smaller grains in comas, but their contribution to the
total dust population remained marginal.
Even if grains smaller than $ 0.1\,\mu$m are produced 
abundantly
in the planetesimal breakup, they are not expected to contribute
significantly to the avalanche process since they are 
in the size range
where $\beta$ decreases for smaller grains (see Fig.~\ref{fig:beta}). 
As a consequence, they 
have lower outgoing velocities which, 
together with their smaller masses,
lead to a marginal contribution in terms of impacting kinetic energy.
We thus believe $0.1\,\mu$m to be a reliable minimum value for 
$s_\mathrm{min,pl}$
and explored $s_\mathrm{min,pl}$ values in 
the $0.1\,\mu$m to $1\,\mu$m range, the latter
value being the one considered by \citet{KenyonBromley05}. 
{
Although the impacting kinetic energy per grain is increasing 
in the $0.1\,\mu$m to $1\,\mu$m size range 
 (leading to an increase of the $N_\mathrm{tot}/N_0$ ratio),
 $F_\mathrm{max} \propto N_\mathrm{tot}/N_0 \times \langle s(t_*)^2 \rangle / \
langle s_0^2 \rangle$ decreases  with increasing
$s_\mathrm{min,pl}$ (Fig.~\ref{fig:fm_sminc})
because of the decreasing value of 
the $\langle s(t_*)^2 \rangle / \langle s_0^2 \rangle$ factor. 
}

Test runs have also been performed to check the $s_\mathrm{max,pl}$ dependence.
This exploration has shown that results do not depend on this parameter
for values higher than 1\,cm. This is due to the fact that 
grains bigger than this
size have very small $\beta$ values, as well as a low
total cross-sectional area, which do not allow them to
significantly contribute to the avalanche propagation.

\begin{figure}[t] 
\begin{center}
\includegraphics[width=\columnwidth,clip]{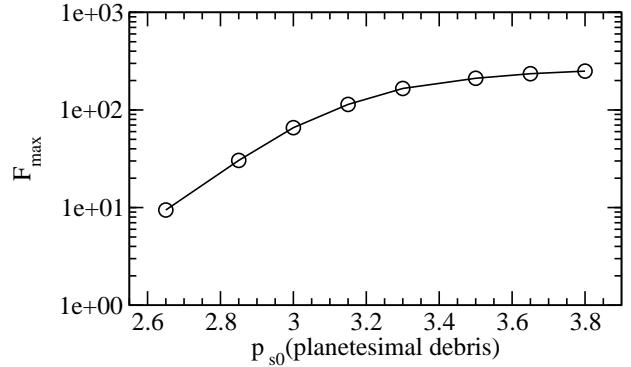}
\end{center}
\caption{Dependence of the maximum amplification factor
on the power-law index of the initial size-frequency 
distribution of the planetesimal debris
 (Eq.~\ref{eq:dn}), $s_\mathrm{min,pl}=0.1\,\mu$m.}
\label{fig:fm_ps0}
\end{figure}

The dependence of $F_\mathrm{max}$ on the  power-law 
index for the initial planetesimal debris size distribution,
 $p_\mathrm{0,pl}$, is shown in  Fig.~\ref{fig:fm_ps0}.
 It can be noted that increasing  $p_\mathrm{0,pl}$ above 3.5 (value for the nominal case)
 does not lead to a significant increase of $F_\mathrm{max}$, whereas
 less steep power laws lead to a significant decrease
  in  $F_\mathrm{max}$.

\subsubsection{Position of the planetesimal breakup} \label{sec:r0c}

\begin{figure}[t]
\begin{center}
\includegraphics[width=\columnwidth,clip]{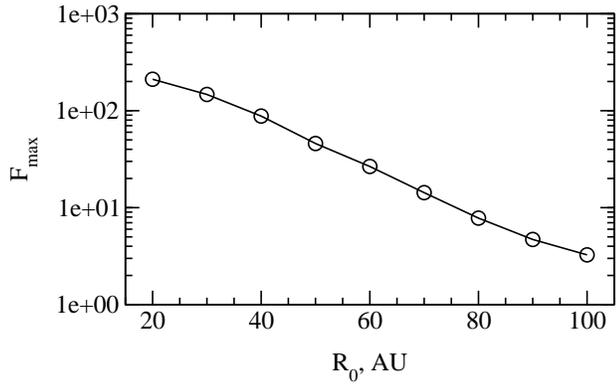}
\end{center}
\caption{Dependence of the maximum \ampl factor on the location of
the primary planetesimal breakup.}
\label{fig:fm_r0}
\end{figure}

\begin{figure}[t]
\begin{center}
\includegraphics[width=\columnwidth,clip]{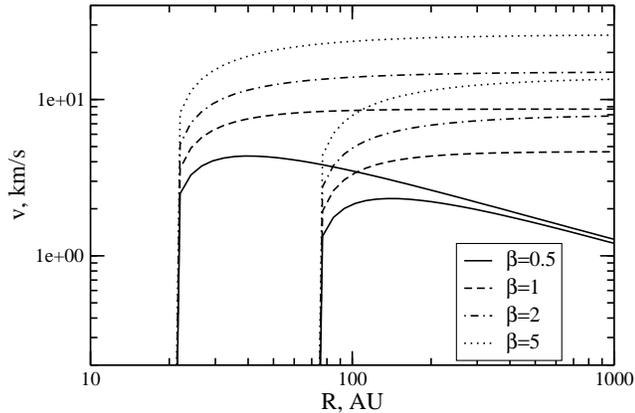}
\end{center}
\caption{Radial velocities vs. distance from the star for grains with
different $\beta$ values
released by parent bodies on circular orbits for  a \bpic-like star.
The release distances
are 20\,AU and 70\,AU. 
}
\label{fig:vr0}
\end{figure}

We perform  a set of runs in which the position of the planetesimal breakup, $R_0$,  
is varied between 20 and 100\,AU (Fig.~\ref{fig:fm_r0}), but
all the other parameters remain identical to the
nominal case.
The maximum amplification factor decreases with increasing $R_0$ for two 
reasons: (i) the total amount of disc material through which 
the outflowing grains propagate is higher when the grains are released
close to the star;
(ii) the unbound grains ($\beta > 0.5$) { have time to}
reach higher radial velocities
if they are released closer to the star (see Fig.~\ref{fig:vr0}), 
which leads to more violent collisions and hence 
higher dust production per collision. 

\subsection{Dependence of $F_\mathrm{max}$ on the prescription for collisional outcome}

\subsubsection{Minimum size of avalanche produced debris}\label{sec:size}

\begin{figure}[t] 
\begin{center}
\includegraphics[width=\columnwidth,clip]{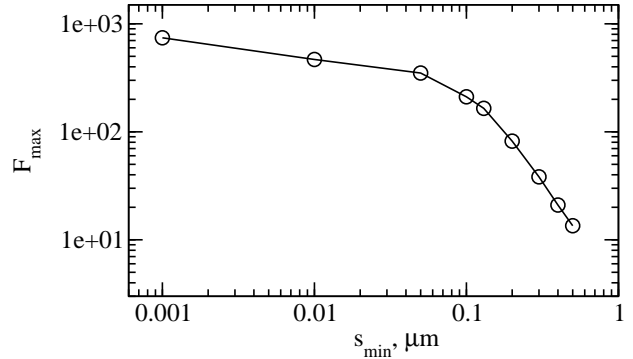}
\end{center}
\caption{Dependence of the maximum amplification factor on the minimum
size for debris produced in avalanche collisions   
(Eq.~\ref{eq:dn_qi} with $q_1=1.5, q_2=1.83, m_s=M_\mathrm{lf}/3$). 
}
\label{fig:fm_smina}
\end{figure}

In the nominal case we assume that the minimum size, $s_\mathrm{min,col}$,
for the debris produced by collisions and the minimum size of the initial
planetesimal debris, $s_\mathrm{min,pl}$, are both equal to $0.1\,\mu$m.
We have seen in Sect.~\ref{sec:sminc} that we do not expect
significant changes when $s_\mathrm{min,pl}<0.1\,\mu$m.  
However, the situation is slightly different for avalanche grains,
since these grains are continuously produced through collisions.
We investigate this parameter's effect in 
test runs exploring different values for $s_\mathrm{min,col}$ (Fig.~\ref{fig:fm_smina}).
As can easily be seen, $F_\mathrm{max}$ does not strongly vary
with $s_\mathrm{min,col}$ for $s_\mathrm{min,col}<0.1\,\mu$m.
There are two reasons for that: i) 
$\beta$  decreases with decreasing sizes for grains smaller then 
$\sim 0.1\,\mu$m (Fig.~\ref{fig:beta}), thus preventing them from
significantly contributing to the avalanche propagation;
ii) the broken power law for the debris size distribution, and
especially the flatter index $q_1=1.5$ for the smallest grains, prevents
them from taking up most of the cross-sectional area of the avalanche grains.

\subsubsection{Size distribution of the debris grains}

\begin{figure}[t] 
\begin{center}
\includegraphics[width=\columnwidth,clip]{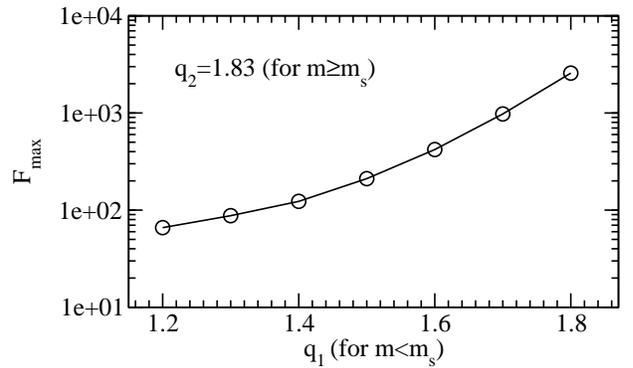}
\end{center}
\caption{Dependence of the maximum amplification factor on  the value 
$q_1$ in Eq.~\ref{eq:dn_qi} for 
  collisionally produced grains
 ($q_2=1.83$). }
\label{fig:fm_q}
\end{figure}

Numerical exploration of the position of the slope
change $m_s$ (with $s_\mathrm{min}=0.1\,\mu$m) and of the $m>m_s$ power-law index $q_2$ 
shows that the resulting amplification factor only weakly depends on 
these parameters. 
Changing  $m_s/M_\mathrm{lf}$ from 1 to 10 leads to changes in
$F_\mathrm{max}$ by only a factor $\sim 2$.  
On the contrary, variation in the $q_1$ index can significantly 
affect $F_\mathrm{max}$, especially
for $q_1>5/3$, when most of the produced cross-sectional area resides in
the smaller grains (Fig.~\ref{fig:fm_q}).

\subsection{Grain chemical composition and impact strength }
\label{sec:qth}

The exact chemical composition of circumstellar 
disc material is not well constrained and might in any case 
vary from one system to the other.
There is observational evidence for silicates, ices, and metals 
\citep[e.g.,][]{Pantin97,Bouwman03}, but their exact proportions in individual
grains are difficult to estimate.  Several detailed studies have
addressed this issue for the specific \bpic~case
 \citep[e.g.,][]{LG98, Pantin97}, but the estimates remain
model dependent.
Changes in grain compositions might affect the results in two ways:
(i) compositional changes can lead to different values of $\beta$, so that
the grains experience different radiation pressure and
as a consequence reach different outgoing (and impacting) velocities, and
(ii) their collisional response properties can be significantly different.

\begin{figure}[t]
\includegraphics[width=\columnwidth,clip]{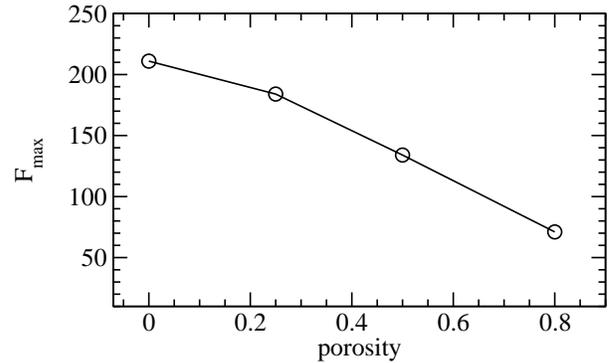}
\caption{Maximum amplification factor as a function of 
porosity for pure silicate grains. The value of the
threshold energy, $Q^*$, is assumed to be the same as in the 
nominal case.}
\label{fig:fmc}
\end{figure}

We first explore the role of grain porosities by varying this parameter
between $P=0$ (compact grains, nominal case) and $P=0.8$ (highly porous grains), 
with $Q^*_0$
remaining constant. This constant $Q^*_0$ prescription might seem counter-intuitive
at first, since more porous grains should be expected to be more
fragile, but it is, in fact, supported by numerical experiments showing
that porous targets often prove more resistant than non-porous ones
\citep[e.g.,][]{FlynnDurda04,Ryan91,Love93}, the reason being that impact shock waves are effectively
dissipated by the pores. Figure~\ref{fig:fmc}  shows that avalanche strength is maximum for the nominal case of
compact grains ($F_{\mathrm{max}(P=0)}\simeq210$) 
and decreases for porous
grains ($F_{\mathrm{max}(P=0.8)}\simeq70$).
  
We numerically explore the importance of chemical composition
by performing runs for the 2 extreme cases of pure (compact) silicates 
(Mg$_\mathrm{0.95}$Fe$_\mathrm{0.05}$SiO$_\mathrm3$) and pure (compact) water ices. 
Here again, we take the possibly
counter-intuitive constant $Q^*_0$ assumption, which is here again
supported by experimental results showing that for target-projectile pairs
of the same material, ices can be as strong as silicates \citep[e.g.,][]{Ryan99}.
Furthermore, compact ices and silicates of equivalent sizes have similar $\beta$ 
values in the $s>0.1\,\mu$m
 range (see Fig.~\ref{fig:beta}). It is thus
not surprising that our results show no significant  difference
 between 
the pure-ice ($F_\mathrm{max}=300$) and 
pure-silicate runs ($F_\mathrm{max}=210$).

\begin{figure}[t] 
\begin{center}
\includegraphics[width=\columnwidth, clip]{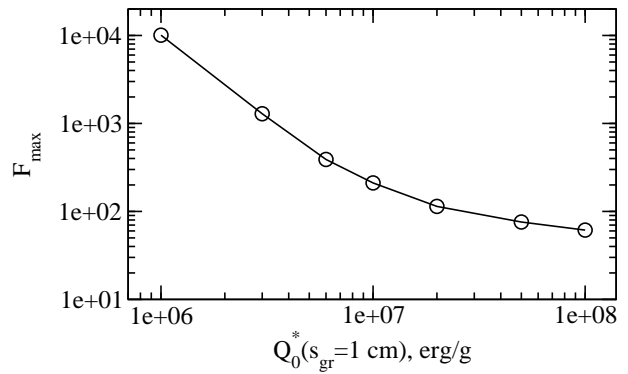}
\end{center}
\caption{Maximum amplification factor vs.
 value of the threshold energy, $Q^*$. Values for $s_0=1$\,cm grains
are denoted on the axis. For the other sizes the threshold energy is
given by Eq.~\ref{eq:Qth1}. }
\label{fig:qth}
\end{figure}

In a third set of runs we separately explore the $Q^*$ parameter, whose
values for given grain compositions and dynamical conditions
are still not well constrained by experiments or theoretical studies.
The threshold energy estimates for silicate-silicate and ice-ice collisions
might vary between $\simeq 10^6$ and a few\,$\times 10^7$\,erg/g
\citep[e.g.,][]{Ryan99,Benz99,Holsapple02}.
We explore $Q_0^*$ values between $10^6$ and $10^8$\,erg/g and
obtain strong variations in $F_\mathrm{max}$ (Fig.~\ref{fig:qth}). For the lowest
explored $Q_0^*$ value of $10^6$\,erg/g, we get 
$F_\mathrm{max}=10^4$, which is about 50 times higher than 
in the nominal case ($Q_0^* = 10^7$\,erg/g).

\subsection{Field particle population} 
\label{sec:fp}

\begin{figure}[t] 
\begin{center}
\includegraphics[width=\columnwidth, clip]{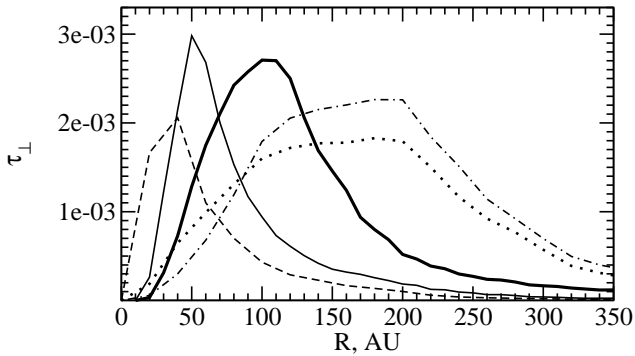}
\end{center}
\caption{Different test radial distributions for $\tau_\mathrm{\perp}$
 ($\tau_{\|}$ being constant). The thick solid line is the 
distribution for the nominal case, taken from \cite{Au01}.
}
\label{fig:tau_r}
\end{figure}

As mentioned earlier, our reference field particle disc
was assumed to be similar to the \bp~system, for which the
dust profile derived by \citet{Au01} has been taken.
 Here we explore alternative profiles
(Fig.~\ref{fig:tau_r}). Results show that 
$F_\mathrm{max}$ does not strongly depend on the shape of
the density  distribution profile as long as the total
radial optical depth of the system (within 500\,AU) $\tau_\mathrm{\|}$
remains the same. 
This result is in agreement with the simplified theory presented in
Sect.~\ref{sec:theory}.

\begin{figure}[t] 
\begin{center}
\includegraphics[width=\columnwidth,clip]{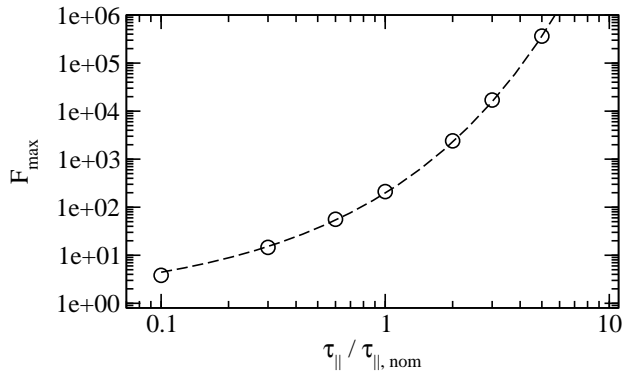}
\end{center}
\caption{Maximum \ampl factor as a function of $\tau_{\|}$.
The open circles are the results of our simulations. The dashed line is the
theoretical prediction (Eq.~\ref{eq:Ftau}).}
\label{fig:tau}
\end{figure}

On the other hand, we get drastic $F_\mathrm{max}$ variations when 
changing the value of $\tau_\mathrm{\|}$ (regardless of the 
radial profile). Figure~\ref{fig:tau}
shows for example that 
increasing the number
      density by a
      factor of 5 leads to a value of  $F_\mathrm{max}$,
       which is a factor of
      $\sim 1000$ higher.
This strong increase in  $F_\mathrm{max}$ is in agreement with
Eq. \ref{eq:FN}, which predicts a strictly exponential growth with
$\tau_\mathrm{\|}$, if $ \langle N_\mathrm{\beta} \rangle $ is constant.
However,
{ in the simulations we find that $\langle N_\mathrm{\beta}\rangle$
weakly varies with $\tau_\mathrm{\|}$ through the
empirical relation $\langle N_\mathrm{\beta}\rangle  \approx 150\,
(\tau_\mathrm{\|}/\tau_\mathrm{\|,nom})^{-0.45}$. 
Plugging this expression for $\langle N_\mathrm{\beta}\rangle $ and  
$ {\langle s(t_*)^2\rangle }/{\langle s_0^2\rangle }\simeq 1.5  $
into Eq.~\ref{eq:FN} we get:
}
\begin{equation}
\begin{array}{ll}
F_\mathrm{max} & \simeq 
\exp\left[240\, \tau_\mathrm{\|} 
\left( \frac{\tau_\mathrm{\|}}{\tau_\mathrm{\|,nom}}\right)^{-0.45}
\right]=  \\      
& = \exp \left[5.3 \left( \frac{\tau_\mathrm{\|}}{\tau_\mathrm{\|,nom}}
 \right)^{0.55}\right] .
\label{eq:Ftau}
\end{array}
\end{equation}
Thus $\tau_\mathrm{\|}$ proves to be the most efficient parameter for increasing
$F_\mathrm{max}$ by several orders of magnitude. This is a point of crucial
importance when considering the absolute strength of an avalanche
and its possible observability.

\section{Avalanche observability} \label{sec:obs}

So far we have been concerned with the way an avalanche 
develops in a disc, in particular with how much dust 
can be created compared to the initial amount of released grains. 
This  was quantified by the amplification factor
$F_\mathrm{max}$.
However, the crucial issue is  under which conditions such
an avalanche might
become {\it observable}. 
In this respect, looking at $F_{max}$ is not enough.
What matters here is the ratio 
between the luminosity of the avalanche particles and 
that of the ``field'' particles, $L_\mathrm{av}/L_\mathrm{d}$.

The value of  $L_\mathrm{av}/L_\mathrm{d}$ corresponding to 
the observability limit
depends on several factors, such as the  physical parameters of the system, 
observational conditions,  and the observing devices' characteristics.
Since our current study is not dedicated to a specific system, 
it is impossible to give a precise criterion for  avalanche observability. 
We shall thus adopt a simple and probably conservative criteria in which
an avalanche is deemed observable when $L_\mathrm{av}/L_\mathrm{d} \ga 1$ 
is reached at a given location in the disc.

{
Most of the resolved debris disc images have been obtained 
in the visual or near-infrared (NIR) domains, 
dominated by scattered starlight.
We shall thus focus here more specifically on scattered
light  luminosity.
The amount of light scattered towards an observer
coming from a given region of the disc is proportional to
\begin{equation}
L\mathrm{}(\lambda_0) \propto 
\int_{V}\!\!\int_{s} F_{*}(\lambda_0,\vec{r})\pi s^2 Q_\mathrm{sca}(s,\lambda_0) 
f(\theta) dn(s,\vec{r}) dV, 
\label{eq:L}
\end{equation}
where the integration is done over the spatial volume of the considered region 
and over the whole
grain size range, $s$, with the grain number density $n$,
the scattering coefficient $ Q_\mathrm{sca}$, and 
the scattering function $ f(\theta) $.
$F_{*}(\lambda_0,\vec{r} )$ is the monochromatic star flux at the distance 
$\vec{r}$ from the star.

For avalanche detection, the visible domain ($\sim\!0.5\,\mu$m) 
is probably more favorable than the NIR ($1-2\,\mu$m).
This is because avalanches consist mostly of submicron grains,
which scatter very inefficiently at $1$--$2\,\mu$m compared to
bigger grains a few microns in size
(which is the average size for the  ``field'' population).
At the same time, in the visual domain
$ Q_\mathrm{sca}$ is nearly the same (within 
a factor of 2, depending on the exact chemical composition)
for  submicron and micron grains. 
Thus the ratio $L_\mathrm{av}/L_\mathrm{d}$ is expected to be higher
in the visual than in the  NIR.
For simplicity we assume that 
$f(\theta) $ is only a function of the scattering angle $\theta$
and that $Q_\mathrm{sca}$ is independent of the grain size.
Although it is not exactly the case, this simplification 
can be considered as a reasonable starting point.
}

Here we consider two extreme cases of disc orientation, i.e., discs
seen exactly edge-on and exactly pole-on. 
{
For the pole-on case, we determine that 
$L_\mathrm{av}/L_\mathrm{d} \simeq$ \ta /\td~and 
 thus use maps of the ratio between these vertical optical
depths.
} For the edge-on case, we
consider the synthetic midplane flux in  scattered light, 
computed for different scattering functions.
Since we do not know the exact optical properties of the circumstellar 
grains, two bracket
cases have been considered for our calculations: isotropic and forward scattering.
For the forward scattering function we use an analytical approximation 
of the
empirical $f_\mathrm{scat}$ for cometary dust, obtained from measurements 
of  solar system comets \citep[][ and references therein]{Art97}:
\begin{eqnarray}
f_\mathrm{scat}(\theta) =f_0 \left[ \frac{0.3}{(0.2+\theta/2)^3} +1.4 
\left(\frac{\theta}{3.3}\right)^4
+0.2\right]  
\label{eq:scat}
\end{eqnarray}

In the following subsections, we investigate under which conditions
avalanche-induced asymmetries might become observable for these two
disc viewing angles.
{ As appears from Figs.~\ref{fig:ass_nominal}-\ref{fig:ass_const},
these asymmetries consist of partial spiral or lumpy patterns
in the face-on case, and of two-sided asymmetries,
for which one side of the system becomes brighter than the other,
in the edge-on case.}
We would like to point out
that at other wavelengths, e.g., infrared, the observability criteria 
($L_\mathrm{av}/L_\mathrm{d}\ga 1$)  might be reached for
lower \ta /\td~ratios due to the fact that { avalanche grains
are expected to be hotter than field particles, since their
average size is about 10 times smaller then the average size of
the initial disc population}.

\subsection{Nominal case} \label{sec:obs_nom}

\begin{figure}[t] 
\makebox[\columnwidth]{
\includegraphics[width=\columnwidth,angle=-90,clip]{rel_nominal_10.ps}
}
\makebox[\columnwidth]{
\includegraphics[width=\columnwidth, clip]{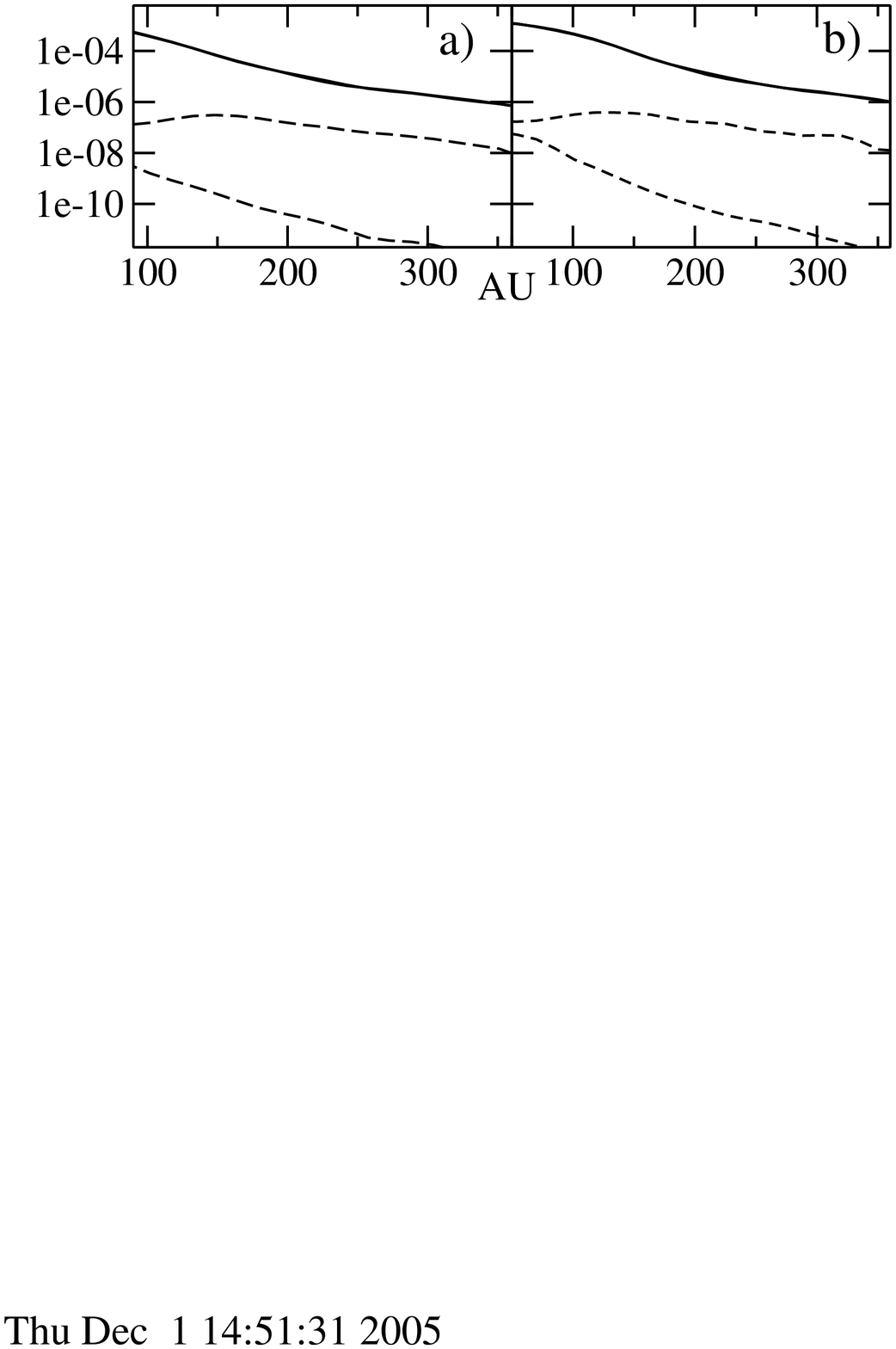}
}
\caption{ {\emph{Top panel:}} 
Face--on case: Color-coded map of the ratio between the 
geometric surface densities of the avalanche grains and 
that of the ``field'' population for the \emph{nominal case}.
{\emph{Bottom panels:}}
Edge--on case: Midplane fluxes (arbitrary units) for the nominal
case at avalanche maximum, edge-on orientation. 
The 2 solid lines indicates the total midplane fluxes
(``field''+avalanche)
{ for each side of the disc (differences between the 2 sides
are so small here that the 2 lines are almost indistinguishable). }  
The dashed lines show the midplane fluxes for just the avalanche particles. 
Plot (a) corresponds to the forward scattering function
and (b) to the isotropic case.}
\label{fig:ass_nominal}
\end{figure}

\begin{figure}[t] 
\makebox[\columnwidth]{
\includegraphics[width=\columnwidth,angle=-90,clip]{weak_grains_16.ps}
}
\makebox[\columnwidth]{
\includegraphics[width=\columnwidth, clip]{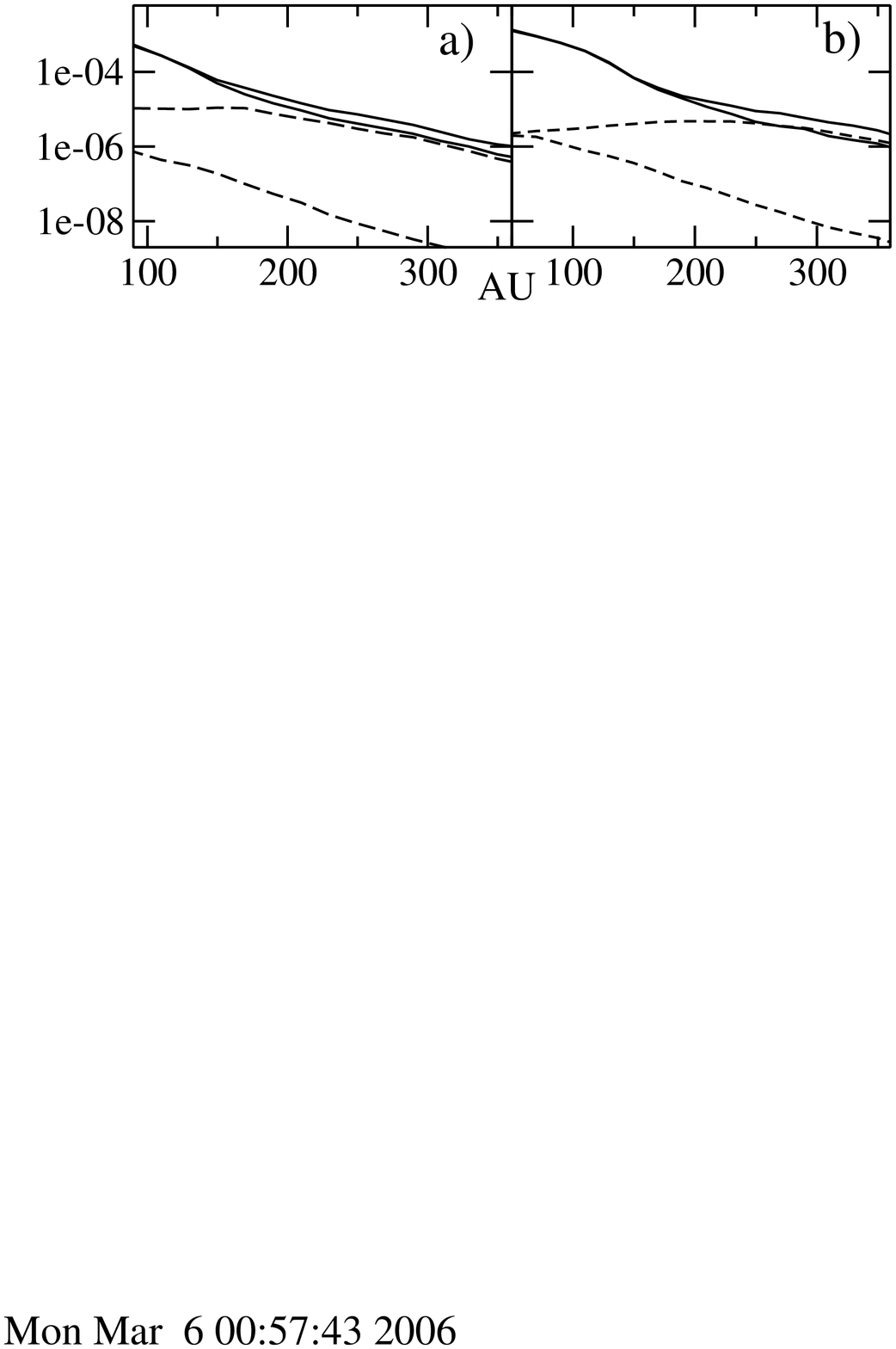}
}
\caption{Same as Fig.~\ref{fig:ass_nominal}, but  
 for \emph{collisionally weaker grains} 
with $Q^*_0(s_0=1\,\mbox{cm})=10^6$\,erg/g (see Sect.~\ref{sec:qth}).}
\label{fig:ass_weak}
\end{figure}

As can be clearly seen in Fig.~\ref{fig:ass_nominal}, 
in the nominal case
$L_\mathrm{av}/L_\mathrm{d}$ never exceeds $10^{-2}$, neither in the edge-on nor
in the head-on configuration. This value is far below our observability 
criterion and the asymmetries induced by the
corresponding avalanche would thus probably be undetectable by scattered
light observations.

\subsection{Larger amount of released dust $M_0$}  \label{sec:mo}

The most straightforward way of getting a more prominent avalanche is to
increase the initially released amount of dust.
As shown in Sect.~\ref{sec:mc},  $F_\mathrm{max}$ remains constant with
varying  $M_0$, so that the ratio $L_\mathrm{av}/L_\mathrm{d}$ 
increases linearly with $M_0$.
As a consequence, the release of $\approx 10^{22}$\,g of dust would be 
required 
for the  avalanche-induced luminosities to  become 
comparable to that of the
rest of the disc.
One might wonder however if a planetesimal shattering releasing
this large amount of dust is a common event (see
discussion in Sect.~\ref{sec:Pobs}).

\subsection{Collisionally weaker grains}

As has been seen in Sect.~\ref{sec:qth}, $F_\mathrm{max}$ increases
strongly for grains with lower specific energy values $Q_*$. 
The lowest $Q_*$ value explored in Fig.~\ref{fig:fmc}, 
$Q^*_\mathrm{0(s_0=1\,\mbox{cm})}=10^6$\,erg/g,  leads to 
\ta/\td $\simeq 0.4$-$0.5$.
Thus, observability might be marginally reached when assuming
the minimum shattering resistance for dust grains.

\subsection{Dustier systems} \label{sec:dd}

\begin{figure}[t] 
\makebox[\columnwidth]{
\includegraphics[width=\columnwidth,angle=-90, clip]{rel_5times_14.ps}
}
\makebox[\columnwidth]{
\includegraphics[width=\columnwidth, clip]{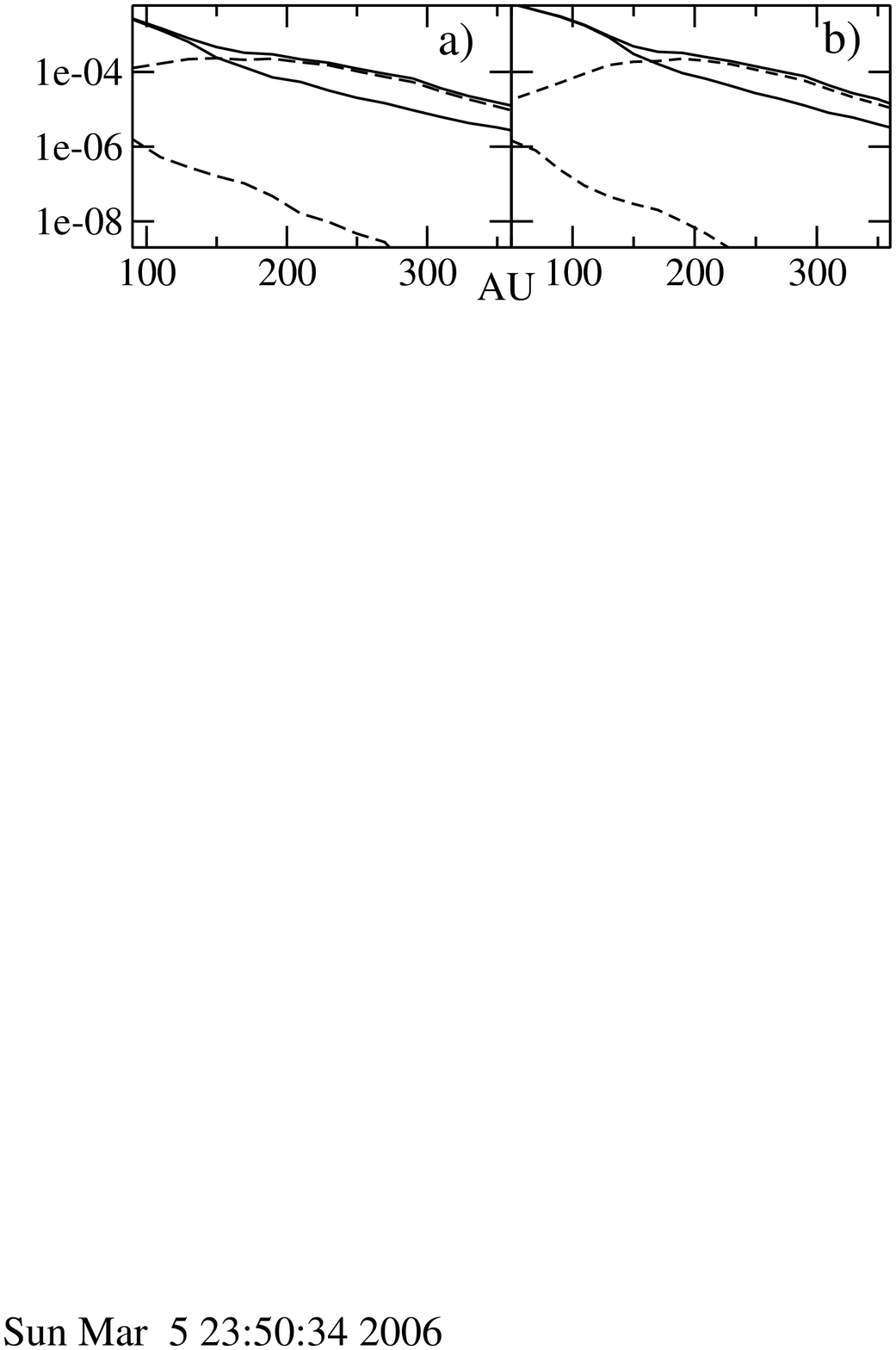}
}
\caption{Same as Fig.~\ref{fig:ass_nominal},
 but  
for the case of \emph{a disc 5 times more massive} than in 
the nominal case.}
\label{fig:ass_5tau}
\end{figure}

The parameter exploration of Sect.~\ref{sec:res} has clearly shown that
the most
efficient parameter for reaching high $F_{max}$ values is
  the field particles number density $\tau_{\|}$
  (Eq.~\ref{eq:Ftau}). 
An obvious way of increasing   $\tau_{\|}$ is to assume a more massive disc,
as has been done in Sect.~\ref{sec:fp}.
In terms of avalanche observability, we find that 
the observability criteria, $L_\mathrm{av}/L_\mathrm{d} \simeq 1$, is
reached for a disc that is  4-5 times more dusty than in the nominal 
case. 
In this case, azimuthal asymmetries become clearly visible in the
face-on configuration  and two-sided brightness
asymmetries for the edge-on case (Fig.~\ref{fig:ass_5tau}).
A massive dusty disc thus looks very promising from the point of 
view of  avalanche
observation.

\begin{figure}[t] 
\makebox[\columnwidth]{
\includegraphics[width=\columnwidth,angle=-90,clip]{thin_disk_15.ps}
}
\makebox[\columnwidth]{
\includegraphics[width=\columnwidth, clip]{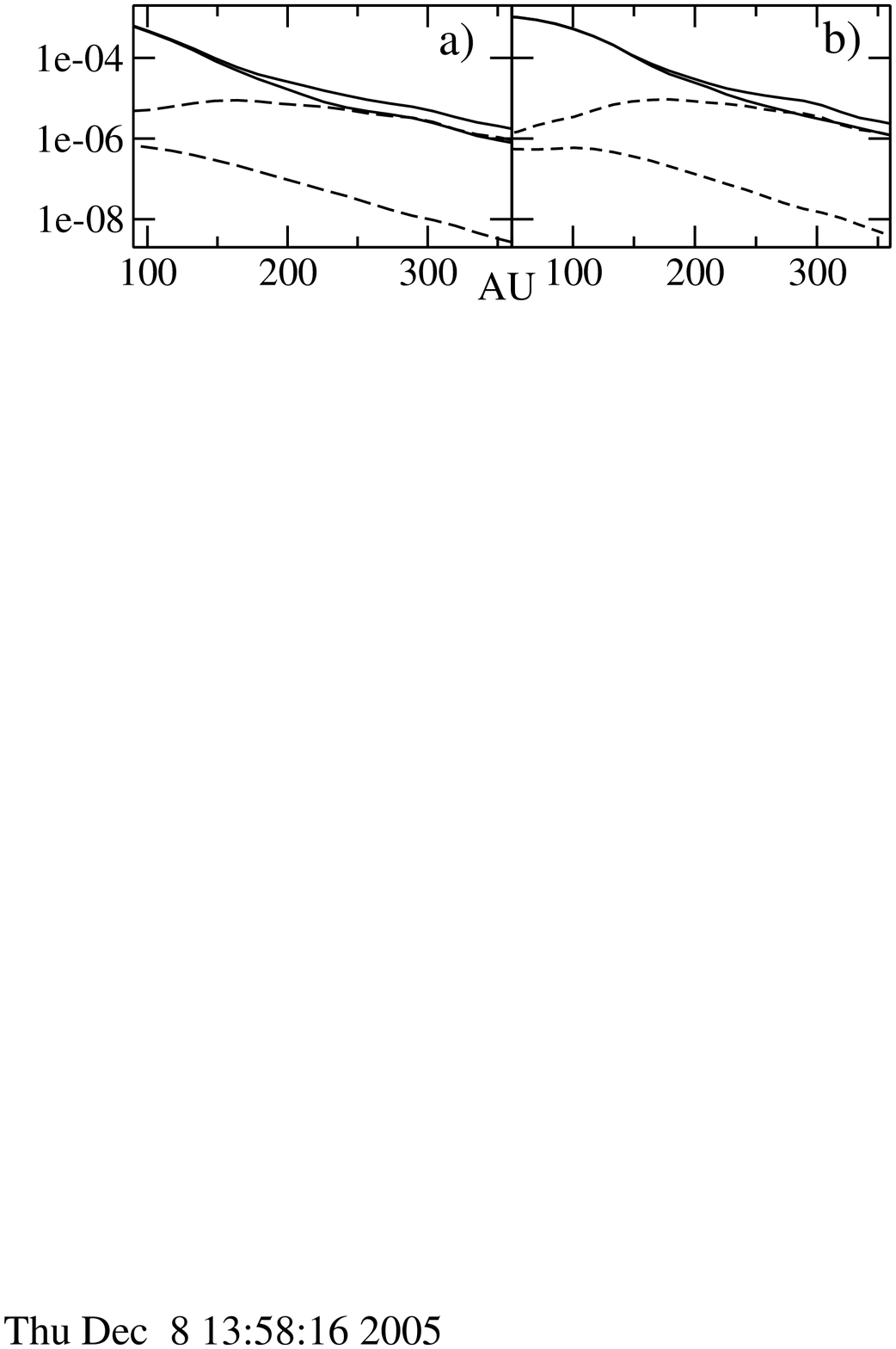}
}
\caption{Same as Fig.~\ref{fig:ass_nominal}, but 
 for \emph{a vertically thinner disc}, described by 
Eq.~\ref{eq:dtauconst}. 
The total mass of the disc is the same  
as in the nominal case, but the asymmetries become prominent.}
\label{fig:ass_const}
\end{figure}

Another way to reach higher values of
$\tau_{\|}$ is to keep the same total amount of dust,
but distributed in  a  vertically thinner disc. 
In the nominal case the dusty disc has the characteristic width $w_\mathrm{}$
(Eq.~\ref{eq:w}), a superexponential vertical profile 
(Eq.~\ref{eq:dtaudz}), and a corresponding $F_\mathrm{max}=210$. 
Assuming now a disc of thickness
 $w^*_\mathrm{}=0.25w_\mathrm{nom}$, 
with a constant vertical profile 
\begin{eqnarray}\label{eq:dtauconst} 
 \frac{d\tau}{dz} = \left\{ \begin{array}{ll}
  \tau(R)/w^*_\mathrm{disc}(R), & \mbox{if $|z|\leq w^*_\mathrm{disc}$} \\
  0, &\mbox{if $|z| > w^*_\mathrm{disc}$,} \\
   \end{array} \right.
\end{eqnarray} 
we get $F_\mathrm{max}=2 \times 10^4$ for the same total amount of dust 
as in the nominal case. In this case,  the number density of
avalanche grains can even largely exceed that of
the field particles (see Fig.~\ref{fig:ass_const}). 
This density enhancement  is azimuthally asymmetric due to
the spiral structure of an avalanche. If the disc is 
orientated  face-on, then 
 the azimuthal asymmetry persists for about 800\,years.
If the system
is viewed edge-on then a two-sided asymmetry can be observed. 
Figure~\ref{fig:ass_const} displays midplane fluxes (``field''+avalanche) 
for different azimuthal angles and scattering functions,
clearly showing that, for favorable system orientations, one side gets significantly 
brighter because of the avalanche.

\section{Discussions} \label{sec:discussion}

\subsection{Probability of witnessing an avalanche event} \label{sec:Pobs}

The  numerical investigation of the previous
sections has shown that collisional avalanches are
a powerful and efficient mechanism that naturally
develops in debris discs after the breakup of a large planetesimal.
However, in our nominal case of a
$\beta$\,Pic-like system and 
$M_0=10^{20}$\,g of dust initially released, the asymmetric
features produced by the avalanche probably remain too weak
to be observable in scattered light (Sect.~\ref{sec:obs_nom}).
This result should, however, be taken with great care
since our parameter exploration has
shown that avalanche strength strongly depends 
on several critical and often poorly constrained parameters.
The first set of parameters is those linked to the initial
breakup event. Here we obtain the intuitive result that
higher amounts of initially released dust leads to 
more powerful avalanches (see Sect.~\ref{sec:mo}),
with the avalanche strength scaling linearly with $M_0$.
This is not unique to the avalanche mechanism: \citet{KenyonBromley05} 
find a similar dependence when only considering the signature of the
cloud of primary debris produced immediately after the planetesimal breakup.
What distinguishes our results from studies
in which only dust released at impact is considered is that avalanches
strongly depend on the number density of  dust in the disc.
Section~\ref{sec:fp} has indeed shown that the global optical depth of
the dust disc $\tau_{\|}$ is the parameter avalanche
development depends most on, the dependence being 
close to an exponential.
We have seen that other parameters, mostly related to the way the
physical response of grains to collisions is modeled, might also
lead to observable events when stretched to the extreme values
that were numerically explored here. This is in particular the case for $Q_*$,
for which very low $\simeq 10^{6}$\,erg/g values might lead to powerful
avalanches.

We shall however leave these ``technical'' parameters aside 
to focus on
the 2 parameters directly related to the system's properties themselves, i.e.,
the optical depth, both $\tau_{\bot}$ and $\tau_{\|}$, and the initial amount
of dust released $M_0$, and  derive   an order-of-magnitude estimate
 for the  probability
of witnessing avalanche events as a function of these parameters.
From the results of Sect.~\ref{sec:res}, the $L_\mathrm{av}/L_\mathrm{d}$
criterion for observability might be written

\begin{equation}
\left( \frac{\tau_{\bot}}{\tau_\mathrm{\bot,nom}} \right)^{-1}
\frac{F_\mathrm{max}}{F_\mathrm{max(nom)}}\,
\frac{M_0}{10^{20}\,\mathrm{g}} \ga 100,
\label{visi0}
\end{equation}  
which is equivalent to saying that 
 the luminosity ratio between 
avalanche and field grains should be at least 100 times
higher than in the nominal case 
(for which $L_\mathrm{av}/L_\mathrm{d} \sim 10^{-2}$). 
Section~\ref{sec:mc} has shown
that $F_\mathrm{max}$ is independent of $M_0$, so that in our approximation
$F_\mathrm{max}$ is only a function of  $\tau_{\|}$, and this  $\tau_{\|}$ 
dependence
is given by Eq.~\ref{eq:Ftau}. 
Thus, Eq.~\ref{visi0} reduces to

\begin{equation}
\exp \left[ 5.3 \left( \frac{\tau_{\|}}{\tau_\mathrm{\|,nom}}
\right)^{0.55}\right] \,
\frac{M_0}{10^{20}\,\mathrm{g}} ~
 \frac{\tau_{\bot,nom}}{\tau_\mathrm{\bot}}\,  \ga  2\times10^{4},
\label{visi1}
\end{equation}  
which gives a direct link between a given disc density ($\tau_{\bot}$ and $\tau_{\|}$)
and the { minimum mass of released dust} able to produce a visible avalanche
in such a disc (the denser the disc, the smaller the corresponding $M_0$ value).
The other important issue affecting witnessing probabilities
is  the duration of
an avalanche. Our simulations show that
the typical lifetime  of 
an avalanche-induced pattern is  $t_\mathrm{av} \sim 10^3$\,yrs.
With this value and Eq.~\ref{visi1}, one can
estimate the probability $P_\mathrm{obs}$
of witnessing an observable avalanche event in 
a given disc:

\begin{equation}
P_\mathrm{obs} = \frac{t_\mathrm{av}}{t_\mathrm{shatt(M_0,\tau_{\bot})}},
\label{probv0}
\end{equation}
where $t_\mathrm{shatt(M_0,\tau)}$ is the average time between
2 shatterings producing $M_0$ of dust in a disc of average
optical depth $\tau_\bot$, with $M_0$,  $\tau_{\|}$ and  $\tau_\bot$ 
satisfying  Eq.~\ref{visi1}. 
{ As suggested in Sect.~\ref{sec:plb}, we consider that the
object releasing $M_0$ of dust has a mass $M_\mathrm{PB}\simeq 10M_0$.}
{
From unpublished results of the \citet{Th03} simulations 
of collisional rates and outcomes in the inner \bpic~disc,
we determine that the approximate timescale for the shattering of
a $M_\mathrm{PB}=10M_0$ object to occur in the innermost $<50$\,AU 
(the typical location
for the initial shattering events considered in our simulations)
of a \bpic~like system is 
$t_\mathrm{shatt}\simeq 150[(10M_0)/10^{21}$\,g$]^{1.25}$\,yrs.
Since, for systems with similar spatial distributions,
the frequency of collisional events is proportional
to the square of a system's total mass, we get the empirical relation:

 \begin{equation}
t_\mathrm{shatt(M_0,\tau_\bot)} \simeq \,150\,
\left( \frac{\tau_\bot}{\tau_\mathrm{\bot,nom}} \right)^{-2}
\left( \frac{M_0}{10^{20}\, \mbox{g}} \right)^{1.25}
\,\mbox{yrs,}
\label{eq:tshat}
\end{equation}
where we implicitly assume that the system's spatial distribution
is the same as in the nominal case, so that the ratio between
two systems' total masses is equal to the ratio
$\tau_\bot / \tau_\mathrm{\bot,nom}$ anywhere in the disc.
This equation should of course be regarded as giving a $very$ rough
estimate, 
since $t_\mathrm{shatt(M_0,\tau_\bot)}$ depends on many poorly
constrained parameters, like the number density of planetesimals and
their average kinetic energy
at impact. 
Equation~\ref{eq:tshat}, however, gives the global trend of the way
$t_\mathrm{shatt(M_0,\tau_\bot)}$ increases with
$M_0$. Taking the lowest $M_0$ value satisfying Eq.~\ref{visi1}
and plugging it into Eq.~\ref{eq:tshat}, 
we get, from Eq.~\ref{probv0}:

\begin{equation}
P_\mathrm{obs}  \approx 3 \times 10^{-5}
\left( \frac{\tau_\bot}{\tau_\mathrm{\bot,nom}} \right)^{0.75}
\exp \left(6.6 \left( \frac{\tau_\|}{\tau_\mathrm{\|,nom}}
\right)^{0.55}\right).
\label{probv2}
\end{equation}
Equation \ref{probv2} indicates that $P_\mathrm{obs}\simeq 0.03$ for the nominal
case field particle disc, which means that we have about a $3\%$ chance of witnessing
the avalanche caused by the breakup of a $M_\mathrm{PB}=10M_0 \sim\!\! 10^{23}$\,g object
($M_0=10^{22}$\,g being the smallest released dust mass able to trigger a visible avalanche
for such a disc, as given by Eq.~\ref{visi1}).
}
This makes it a rather unlikely event, although it cannot be completely ruled out.
Nevertheless, slightly
denser discs (i.e., higher $\tau_{\bot}$ and $\tau_{\|}$) can easily
raise  $P_\mathrm{obs}$ up to 1. As a matter of fact, the dependence on $\tau_{\|}$
is so sharp that $P_\mathrm{obs}=1$ is obtained for
$\tau_{\|}\simeq 2.1\tau_{\|,nom} \simeq  0.046$.
We thus see that a $\beta$\,Pic-like system is below, but not
too far from the limit for which chances of witnessing an avalanche
are high, especially when considering the uncertainties regarding
avalanche strength due to its dependence on several poorly constrained
parameters related to the collision-outcome prescription
(also keeping in mind that higher $\tau_{\|}$ values could alternatively
be achieved for a thinner disc of the same dust mass (see Sect.~\ref{sec:dd})).
{ Moreover, our $L_\mathrm{av}/L_\mathrm{d}>1$ criterion for
observability is probably too conservative, and avalanche-induced
patterns might be detectable for lower luminosity excess values.
Taking, for example, $L_\mathrm{av}/L_\mathrm{d}>0.1$ would raise the
detection probability to $\simeq 45\%$ for a $\beta$\,Pic-like system}.

{

\subsection{Avalanches in observed systems, perspectives}

We defer a detailed application of our model to specific circumstellar discs
to a future study. However, the present results can already give
a good idea of the typical profile for a ``good'' avalanche-system candidate

Our numerical exploration has shown that structures that are the most
likely to be associated with avalanche-events have
two-sided asymmetry for discs viewed edge-on and
open spiral patterns for discs viewed pole-on 
or at intermediate inclinations. An additional requirement is that
these discs should be dust-rich systems, with a dustiness
at least equal to, and preferably higher than that of $\beta$-Pic.
Note also that our model makes an additional prediction,
i.e., that avalanche affected regions should consist of
grains significantly smaller than the "field" particles in
the rest of the disc. If the blow-out radius of grains
is of the order of the wavelength of the observed light,
then this should translate into color
differences between avalanche (bluer) and non-avalanche
(redder) regions.

In this respect, one good edge-on candidate might be the recently discovered
HD\,32297 system, which
exhibits a strong two-sided asymmetry.
As reported by \citet{Schneider05} and \citet{Kalas05}, this system is a $\beta$\,Pic
analog with its SW side significantly brighter than
the NE one within $\simeq$ 100\,AU \citep{Schneider05}
and possibly outside 500\,AU \citep{Kalas05}.
Such a two-sided asymmetry would be compatible
with the ones obtained in our simulations (as shown for example
in the bottom panel of Fig.~\ref{fig:ass_const}). 
Furthermore, \citet{Kalas05} also reported a color asymmetry
between the two sides, with the brighter one (SW) being
significantly bluer. This seems to indicate that this side
is made of smaller, possibly submicron grains \citep{Kalas05}.
As previously discussed, this is
 what should be expected for an avalanche-affected region.
However, an alternative scenario, like the collision with a
clump of interstellar medium proposed by \citep{Kalas05}, might
also explain the HD\,32287 disc structure. Future imaging
and spectroscopic observations are probably needed before
reaching any definitive conclusions.

Among all head-on observed systems, the one displaying the most
avalanche-like structure is without doubt
HD\,141569 \citep{ClampinKrist03}, with its pronounced spiral
pattern. Furthermore, the disc's mass, significantly
higher than $\beta$-Pictoris, makes it a perfect candidate in terms
of witnessing probabilities. Of course, avalanche is
not the only possible scenario here, and several alternative
explanations, like an eccentric bound planet or 
stellar companion, or a stellar flyby  
have already been proposed \citep[e.g.,][]{Au04, Wyatt05, Ardila05}.
One should, however, be aware that
this system is strictly speaking not a "standard" debris disc
as defined by \citet{Lagrange00} and as considered in the present study.
Indeed, several studies seem to suggest the presence
of large amounts of primordial gas \citep{Zuckerman95,Ardila05}.

Gas drag effects have been left out of the present study on purpose,
mainly because, in the strict sense of the term,
debris discs are systems where dust dynamics is 
not dominated by gas friction \citep{Lagrange00}. 
Moreover, the correct description of dust-gas coupling adds several
additional free parameters (gas density and temperature distributions,
etc.) and requires a full 2-D or 3-D treatment of gas
by far exceeding the scope of the present paper.
However, the issue of avalanches in a gaseous medium might
be a crucial one for those systems that are most favorable
for avalanches, i.e., discs more dusty than
$\beta$-Pic, a system which is already at the upper end
of debris-discs in terms of dustiness \citep[e.g.,][]{Spangler01}.
Such more massive systems should fall into a loosely defined
category of "transition" discs between T-Tauri or Herbig Ae
protoplanetary systems and "proper" debris discs 
\citep[see for example Sect.~4 of][]{Dutrey04}.
For such systems (of which HD141569 is a typical example), 
which are younger than the more evolved 
debris discs, risks (or chances) of encountering
large amounts of remaining gas are high.
This crucial issue will be the subject of a forthcoming
paper.}

\section{Summary} \label{sec:sum}

This paper presents the first quantitative study of the
collisional avalanche process in debris discs, i.e.,
the chain reaction of dust grain collisions triggered by the
initial breakup of a planetesimal-like body.
We have developed  a code that
allows us to  simultaneously  follow both spatial \emph{and}  size
distributions of the dust grain population, which is collisionally
evolving because of impacts caused by small particles  blown out of
the system by the star's radiation pressure. Our results 
can be summarized as follows: 

\begin{enumerate}
\item 
Collisional avalanches propagate outwards
{
leaving a characteristic spiral-shaped pattern in the system.
Depending on the system's orientation, these patterns
might appear as open spirals or lumpy structures
(face-on geometry) or a two-sided brightness asymmetry
(edge-on case).
}
In a $\beta$\,Pic-like disc, an avalanche lasts for about $10^3$\,years. 
\item
The strength  of an avalanche depends linearly on the mass
of the initially shattered object, but nearly exponentially
on the optical depth of the dust disc in which it propagates.
The disc's dustiness is by far the most crucial parameter here,
making dusty discs much more favorable cases for avalanche propagation
than tenuous ones.
\item
We define a conservative criterion for avalanche observability, from which
we infer a relation between a given disc density and the minimum mass
of the object that has to be shattered to reach observability.
When coupling this relation to estimates for catastrophic
disruption probabilities among planetesimal-objects in debris discs,
we are able to derive a first-order estimate for the probability of
witnessing an observable avalanche event in a given debris disc.
For our reference $\beta$\,Pic-like system it is
of a few percents, but probabilities rapidly increase for 
slightly denser systems.
\item
Modeling of dustier young transitional discs may require the inclusion of gas drag,
which may change both the morphology and the strength of the avalanche.

\end{enumerate}

\appendix{}
\section{Superparticles' structure} \label{app:sp}

All SPs are  cylinders with 
variable height \hsp and constant radius \rsp. Their geometrical centers 
always stay in the disc midplane. 
For most of the runs we take $r_\mathrm{sp}=5$\,AU.
Test runs showed that an avalanche development  weakly depends on the
value for $r_\mathrm{sp}$. The maximum amplification factor
for runs with \rsp in the $4-8\,$\,AU range
differs by less then $\sim 10\%$. 
SPs are modeled  in different ways depending on 
the physical origin
of the grains they represent. We distinguish between 3 types of SP.

\subsection{Field SP}

All SPs that represent the initial structure of the dusty disc
(i.e., non-avalanche SPs)
have a superexponential vertical density profile
\begin{equation} \label{eq:nz}
 n_{{\mathrm{gr}},i}= \frac{N_{{\mathrm{gr}},i}}
 {\pi r_{{\mathrm{sp}},i}^2 
 \int_{-0.5h_{\mathrm{sp},i}}^{0.5h_{\mathrm{sp},i}} f(z) dz} 
 f(z),
\end{equation}  
where $N_{{\mathrm{gr}},i}$ is the total number of grains in a given
 SP, $r_\mathrm{sp} and h_\mathrm{sp}$
are the SP's radius and height, and $f(z) $ 
is the SP profile function, which reads 
\begin{equation} \label{eq:fz}
f(z)=\exp 
\left( - \left(\frac{|z|}{w_{{\mathrm{sp}},i}(R)}\right)^{p_\mathrm{z}}\right),
\end{equation} 
where  $p_\mathrm{z}=0.7$ and $w_{{\mathrm{sp}},i}$ is the SP's width, which 
depends  on the 
distance to the star, $R$, and is equal to the disc width (Eq.~\ref{eq:w}) 
for field SPs.
The height of a \emph{field} SP is
\begin{eqnarray} 
 h_{{\mathrm{sp}},i}(R)=20w_{{\mathrm{sp}},i}(R). 
\end{eqnarray}  

All field SPs  standing for the biggest grains (1\,cm) have
circular orbits. The SPs radial distribution and the 
 number of grains in each SP is chosen in accordance with the 
 best-fit parent body distribution from \citet{Au01}. All other grains
 are assumed to be produced from these biggest grains  following the
 power law size-frequency distribution of Eq.~\ref{eq:dn}. 
 The number of SPs in each size bin is taken  such that
 in the steady-state configuration 
 there are 2-5 overlapping SPs of the same grain size at any given location in the system.

\subsection{Initially released planetesimal debris} 

When a planetesimal  is shattered the fragments are created with
an initial spread in velocities. 
\cite{RyanMelosh98} show that the fragment velocities depend on grain sizes
and that for small
grains they are about a few percent of the impact velocity, with
$v_\mathrm{fr} \approx 0.01-0.1 v_\mathrm{impact}$. 
Fragments produced from a body on a Keplerian orbit 
would spread in the vertical direction, and the height of the layer can be estimated as
\begin{eqnarray} 
h_\mathrm{sp}\approx 2 \frac{v_\mathrm{fr}}{v_\mathrm{impact}} R.
\end{eqnarray} 
 For most of the runs we take $v_\mathrm{fr} /v_\mathrm{impact}=0.1$. 

If the debris have velocities with isotropic distributions around the center of mass then
the vertical distribution of grains of the same $\beta$ can be approximated
as a constant.
Thus these SPs  are assumed to have no internal density structure and constant 
grain density. 
\begin{eqnarray} 
 n_{{\mathrm{gr}},i}=\frac{N_{{\mathrm{gr}},i}}{\pi r_{{\mathrm{sp}},i}^2 h_{{\mathrm{sp}},i}},
\label{eq:ncomet}
\end{eqnarray} 
which corresponds to 
 $f(z)= 1$ (i.e., $w_{{\mathrm{sp}},i}= \infty$ in Eq. ~\ref{eq:nz}).

\subsection{SP created due to collisions} 
\label{app:new_sp}

As mentioned in Sect.~\ref{sec:model},
when two SPs are passing through each other, a fraction of their grains 
can be destroyed producing new and smaller grains, which are
then combined into new SPs.  
The number of grains that are destroyed in each SP, if the relative velocity is high
enough to reach catastrophic fragmentation (see Sect.~\ref{sec:col}),
is calculated as
\begin{eqnarray} 
\label{eq:npar} 
 N_\mathrm{gr, -}=  \int_0^{t_\mathrm{col}}\!\!\int_\mathrm{-z_\mathrm{0}}^{z_\mathrm{0}}
\sigma n_{{\mathrm{gr}},i} n_{{\mathrm{gr}},j}v_\mathrm{rel}
 A_\mathrm{over}dz dt,
\end{eqnarray}
where $z_\mathrm{0}=0.5\,\mbox{min}(h_{{\mathrm{sp}},i},h_{{\mathrm{sp}},j})$, 
$\sigma=\pi (s_{{\mathrm{gr}},i} +s_{{\mathrm{gr}},j})^2/4$ is 
the collisional cross-Sect.  for grains with physical sizes $s_{{\mathrm{gr}}(i,j)}$, 
$v_\mathrm{rel}$ is the relative velocity of the grains, $A_\mathrm{over}$
is the overlapping area of the two SPs viewed top-on, and $t_\mathrm{col}$
is the time while the SPs are passing through each other.
The time dependences in  
Eq.~\ref{eq:npar} might be neglected if a reasonably small ``collisional''
time step is taken for the calculations. We adopt
$\Delta t_\mathrm{col}=0.02$ (in units of the orbital period at 20\,AU)
for most of the runs.
Simulations with much smaller collisional time steps
(e.g., $\Delta t_\mathrm{col}=0.005$) do not lead
to a significant improvement of the results. 
The debris velocities and the size-frequency distribution for the newly created grains
are identical to the values
that one would get, considering a collision between $N_\mathrm{gr, -}$ pairs of 
grains with  sizes $s_\mathrm{gr,(i,j)}$ and relative velocity $v_\mathrm{rel}$
(Sect.~\ref{sec:col}). 

The structure of the new SPs (one SP for each size bin),
which are produced after a collision, is obtained through the following equations.
The initial height of the SP is equal to 
\begin{eqnarray} 
 h_{{\mathrm{sp}},k}= min(h_{{\mathrm{sp}},i},h_{{\mathrm{sp}},j}).
\end{eqnarray}
The SP vertical profile $f_k(z)$ is calculated as
\begin{eqnarray} 
 f_k(z)=f_i(z) f_j(z), 
\end{eqnarray}
which  corresponds to a SP of width  (Eq.~\ref{eq:fz}) 
\begin{eqnarray} 
 w_{{\mathrm{sp}},k}=\left[ \left(\frac{1}{w_{{\mathrm{sp}},i}}\right)^{-p_\mathrm{z}} + 
 \left(\frac{1}{w_{{\mathrm{sp}},j}}\right)^{-p_\mathrm{z}} \right]^{-1/p_\mathrm{z}}.  
\end{eqnarray}

In the case of dust production by SPs with superexponential profiles
(Eq.~\ref{eq:fz}), the new SPs will have a smaller width
than the two ``colliding'' SPs. Keeping the same width would allow us 
to overlook the fact
that ``real'' grains in the SPs have vertical
velocities and so will the produced debris.
This component
can be neglected if we only consider collisional outcomes, but it should be
taken into account for
calculations of the density structures. In our calculations this effect is taken into
account by increasing the newly-created-SPs' width and height with time.
The increase rate is taken to be
 \begin{eqnarray} 
 \dot{w}_\mathrm{sp}=\frac{w_\mathrm{disc}-w_\mathrm{sp}}{\Delta t_w}, 
\end{eqnarray} 
where $\Delta t_w$ is equal to $\frac{1}{4}$ of the orbital period at the location
where the SP is created.
The same kind of time dependence is applied to the \hsp growth, namely
\begin{eqnarray} 
 \dot{h}_\mathrm{sp}=\left\{ \begin{array}{ll}
 \frac{h_\mathrm{max}-h_\mathrm{sp}}{\Delta t_h} &\mbox{if  $h_\mathrm{sp} < h_\mathrm{max}$} \\
 0  &\mbox{if $ h_\mathrm{sp} \geq h_\mathrm{max},$}
 \end{array} \right.
\end{eqnarray} 
where $ h_\mathrm{max}=h_0 w_\mathrm{disc}$, $t_h=h_0 t_w$, and $h_0$ is a
 constant parameter. The results 
for $h_0=0.3$ and $h_0=3$ differ by a factor of 2, which is acceptable for our
order-of-magnitude calculations.

\section{Recombining SPs}

To keep the total number of SPs manageable (we can  trace 
the evolution of about one million of them) we have developed an
algorithm that allows us to recombine
SPs with similar parameters (grain size, velocities, positions in the
disc). This allows us to avoid a too fast increase of the total
number of SPs while keeping it to a value
large enough from a statistical point of view.

The merging procedure is applied 
only between SPs standing for grains of \emph{the same size}.
The proximity condition is obtained by dividing the disc plane into
a 2-D space grid with the cell size
equal to the SP's diameter. For each cell we list all SPs whose
centers are located within the cell. 
Thus a given list contains SPs from the same spatial 
volume and  with the same grain size.  
For each SP from a given list we calculate the ``normalized'' 
velocity $v^*=v/v_\mathrm{kep}(R)$, where
$v$ is the SP velocity and $R$ is the distance from the star to the
SP center. If several SPs fall into 
\emph{one velocity bin} (of width $dv^*=10^{-2}$), 
they are combined into one SPs with the number of grains being equal 
to the sum of $N_\mathrm{gr}$ of the combined SPs. 
The vertical structure of the new SP is 
obtained through the averaged values 
$w_{\mathrm{sp}}=\frac{\sum w_{\mathrm{sp},i} N_{\mathrm{gr},i}}
{\sum  N_{\mathrm{gr},i}}$ and
$ h_{\mathrm{sp}}=\frac{\sum h_{\mathrm{sp},i} N_{\mathrm{gr},i}}
{\sum  N_{\mathrm{gr},i}}$.
The velocity and the position of the new SP are 
chosen to be equal to those of one SP, randomly chosen from
the list of the  recombined SPs. This SP is chosen randomly with
the probability proportional to $\frac{ N_{\mathrm{gr},i}}
{\sum  N_{\mathrm{gr},i}}$. Choosing positions and velocities this way
has the advantage of not introducing new trajectories for SPs,
and this  makes us more certain about the general shape of an
avalanche. We have also tested another procedure for SPs recombination
in which all SPs from the same velocity bin are recombined into one
 SP with  position and velocity such that the total
  angular momentum and 
kinetic energy are preserved. In both cases we obtain similar
results.

{
The recombination described above makes a significant reduction
of  the total number of avalanche SP spossible. However,
an additional optimization procedure has been implemented
 to speed up the calculations.
The idea is the following: The dust 
production rate is  approximately proportional to the number density 
of avalanche grains (multiplied by the number
 density of the field population). 
As a consequence, regions with the lowest  density of avalanche grains 
(of a given size)  cannot make a significant
contribution to avalanche dust production. 
Thus these regions are not very interesting
 for our simulations 
and there is no need to do very accurate representations of the avalanche 
grains' population there. A very rough representation is enough here.
The question is to determine which number density values
should be considered as "unproductive".
Since midplane densities for the field population
vary by a factor $\sim\!100$ throughout the disc,
regions where the density of
 avalanche grains (of a given size bin) are less than $1/1000$
that of the most dense regions will be considered as "not important".
In these regions all SPs from the same space cell 
(the proximity condition as defined above) representing grains of
 the same size bin are combined into one SP regardless of their velocities.
Test runs have proved the efficiency of this optimization procedure:
i) the general shape of an avalanche and the amplification factor evolution are 
not modified by this procedure, since the "important" regions are not affected;
 ii) the total number of SPs is significantly reduced, significantly
speeding up the calculations. However, this procedure is limited  to 
cases for which there is a significant density gradient. 
At later stages of avalanche propagation ($\ga 12$ orbital periods) 
it is inefficient. Figure~\ref{fig:nsp} shows the evolution of the 
total number of SPs with time.

\begin{figure}[t] 
\begin{center}
\includegraphics[width=\columnwidth, clip]{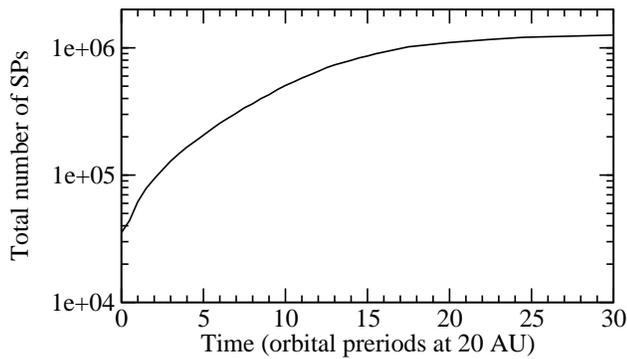}
\end{center}
\caption{The total number of SPs as a function of time for
a typical nominal case run. 
}
\label{fig:nsp}
\end{figure}

\

}

\bibliographystyle{aa}
\bibliography{bibliography}

\end{document}